\documentclass[12pt]{article}
\pdfoutput=1

\usepackage{cite}
\usepackage{booktabs}
\usepackage[english]{babel}
\usepackage{amsmath,amssymb,amsbsy,amstext, amsthm, simplewick}
\usepackage{hyperref}
\usepackage{graphicx}
\usepackage{amsfonts}
\usepackage{amssymb}
\usepackage[small]{caption}
\usepackage{upgreek}
\usepackage[svgnames,dvipsnames,x11names,table]{xcolor}
\usepackage{multirow}
\usepackage{geometry}
\usepackage[hang,flushmargin]{footmisc}
\usepackage{bm}
\usepackage{braket}
\usepackage{subcaption}
\usepackage{mathtools}
\usepackage{setspace}
\usepackage{cleveref}
\usepackage{comment}
\usepackage{scalerel}
\usepackage[normalem]{ulem}
\usepackage{slashed}
\usepackage{enumitem}
\usepackage{dsfont}
\usepackage{tikz}
\usetikzlibrary{decorations.markings}
\usetikzlibrary{shapes.misc}
\usetikzlibrary{arrows}
\usetikzlibrary{spy}
\usetikzlibrary{calc}
\usepackage{pgfplots}

\makeatletter
\g@addto@macro\bfseries{\boldmath}
\makeatother

\hypersetup{
    colorlinks=true,
    linkcolor={red!50!black},
    citecolor={blue!50!black},
    urlcolor={blue!80!black}
}

\usepackage{colortbl}

\makeatletter
\newlength{\apb@width}
\newcommand{\autoparbox}[2][c]{\settowidth{\apb@width}{#2}\parbox[#1]{\apb@width}{#2}}

\makeatother

\definecolor{lightgray}{gray}{0.9}

\usepackage[framemethod=default]{mdframed}
\newmdenv[skipabove=7pt,
skipbelow=7pt,
rightline=false,
leftline=false,
topline=false,
bottomline=false,
backgroundcolor=gray!10,
linecolor=gray,
innerleftmargin=5pt,
innerrightmargin=5pt,
innertopmargin=5pt,
innerbottommargin=5pt,
leftmargin=0cm,
rightmargin=0cm,
linewidth=4pt]{eBox}

\usepackage[most]{tcolorbox}
\tcbset{colback=white, colframe=black,
        highlight math style= {enhanced, 
            colframe=red,colback=red!10!white,boxsep=0pt}
        }
\definecolor{light-gray}{gray}{0.95}

\crefname{table}{Table}{Tables}
\crefname{equation}{Eq.}{Eqs.}
\crefname{appendix}{App.}{Apps.}
\crefname{section}{Sec.}{Secs.}
\crefname{figure}{Fig.}{Figs.}

\numberwithin{equation}{section}

\def\beq{\begin{equation}}
\def\eeq{\end{equation}}

\def\bea{\begin{eqnarray}}
\def\eea{\end{eqnarray}}

\def\vp{\varphi_{+}}
\def\vm{\varphi_{-}}
\def\dvp{{\dot \varphi}_{+}}

\def\bvp{{\bar \varphi}_{+}}

\def\d{{\rm d}}

\def\beq{\begin{equation}}
\def\eeq{\end{equation}}
\def\bea{\begin{eqnarray}}
\def\eea{\end{eqnarray}}

\def\L{{\cal L}}

\def\d{{\rm d}}

\def\deq{{:=}}

\def\d{{\rm d}}

\def\k{{\vec{\scaleto{k}{7pt}}}}

\def\x{{\vec x}}

\def\X{{\widetilde{X}}}

\def\t{\texttt{t}}

\newcommand{\iid}{i.i.d.}

\DeclareRobustCommand{\SkipTocEntry}[4]{}

\newcommand{\s}{\hspace{0.8pt}}

\definecolor{colorTC}{rgb}{.2,.7,.2}

\definecolor{amethyst}{rgb}{0.6, 0.4, 0.8}

\definecolor{acolor}{rgb}{0.4, 0.2, 0.4}

\definecolor{blue3}{RGB}{31, 119, 180}
\definecolor{red3}{RGB}{	214, 39, 40}
\definecolor{orange3}{RGB}{255, 127, 14}
\definecolor{green3}{RGB}{44, 160, 44}

\begin{document}

\begin{titlepage}
\setcounter{page}{1} \baselineskip=15.5pt
\thispagestyle{empty}
$\quad$
{\raggedleft CERN-TH-2022-206\par}
\vskip 60 pt

\begin{center}
{\fontsize{18}{18} \bf Large Deviations in the Early Universe}
\end{center}

\vskip 20pt
\begin{center}
\noindent
{\fontsize{12}{18}\selectfont  Timothy Cohen$^{\s 1,2,3}$, Daniel Green$^{\s 4}$, and Akhil Premkumar$^{\s 5}$}
\end{center}

\begin{center}
\vskip 4pt
\textit{ $^1${\small Theoretical Physics Department, CERN, 1211 Geneva, Switzerland}
}
\vskip 4pt
\textit{ $^2${\small Theoretical Particle Physics Laboratory, EPFL, 1015 Lausanne, Switzerland}
}
\vskip 4pt
\textit{ $^3${\small Institute for Fundamental Science, University of Oregon, Eugene, OR 97403, USA}
}
\vskip 4pt
\textit{ $^4${\small Department of Physics, University of California at San Diego,  La Jolla, CA 92093, USA}
}
\vskip 4pt
\textit{ $^5${\small Kavli Institute for Cosmological Physics, University of Chicago, IL 60637, USA}
}

\end{center}

\vspace{0.4cm}
 \begin{center}{\bf Abstract}
 \end{center}

\noindent
Fluctuations play a critical role in cosmology.
They are relevant across a range of phenomena from the dynamics of inflation to the formation of structure.
In many cases, these fluctuations are coarse grained and follow a Gaussian distribution as a consequence of the Central Limit Theorem.
Yet, some classes of observables are dominated by rare fluctuations and are sensitive to the details of the underlying microphysics.
In this paper, we argue that the Large Deviation Principle can be used to diagnose when
one must to appeal to the fundamental description.
Concretely, we investigate the regime of validity for the Fokker-Planck equation that governs Stochastic Inflation.
For typical fluctuations, this framework leads to the central limit-type behavior expected of a random walk.  However, fluctuations in the regime of the Large Deviation Principle are determined by instanton-like saddle points accompanied by a new energy scale. When this energy scale is above the UV cutoff of the EFT, the tail is only calculable in the microscopic description.
We explicitly demonstrate this phenomenon in the context of determining the phase transition to eternal inflation, the distribution of scalar field fluctuations in de Sitter, and the production of primordial black holes.

\end{titlepage}

\setcounter{page}{2}

\restoregeometry

\begin{spacing}{1.2}
\newpage
\setcounter{tocdepth}{2}
\tableofcontents
\end{spacing}

\setstretch{1.1}

\section{Introduction}

The cosmological evolution of our Universe was shaped by fluctuations. The formation of dark matter halos, and hence galaxies and galaxy clusters, was the result of large density fluctuations, which can be modeled using Gaussian random fields. Rare fluctuations, which are determined by the tail of the probability distribution, may also be important for cosmology.  For example, determining if some (or all) of the abundance of dark matter is due to the presence of primordial black holes requires a precise knowledge of the tail of the distribution. Such rare fluctuations could also have played a critical role in the dynamics of (eternal) inflation at very early times.

Despite the broad interest in this subject (\emph{e.g.}~for inflation see~\cite{Chen:2006nt,Seery:2007we,Leblond:2008gg,Shandera:2008ai,Burgess:2009bs,Baumann:2011su,Baumann:2011ws,Flauger:2013hra,Baumann:2014cja,Baumann:2015nta,Adshead:2017srh,Babic:2019ify,Grall:2020tqc} and for eternal inflation see~\cite{Bousso:2006ge,ArkaniHamed:2007ky,Seery:2009hs,Matsui:2018bsy,Kinney:2018kew,Brahma:2019iyy}), characterizing the regime of validity of perturbation theory for the tail of the distribution is an under developed subject. The probability distribution of a field in a de Sitter background can be calculated using a Fokker-Planck equation, known as the framework of Stochastic Inflation~\cite{Vilenkin:1983xq,Starobinsky:1986fx} (see also \cite{Aryal:1987vn,Nambu:1987ef,Kandrup:1988sc,Salopek:1990jq,Starobinsky:1994bd,Wands:2000dp,Tsamis:2005hd,Seery:2005gb,Wands:2010af,Finelli:2010sh,Tada:2016pmk,Achucarro:2016fby,Markkanen:2019kpv,Baumgart:2019clc,Baumgart:2020oby,Pinol:2020cdp}).  Recently, the introduction of the Soft de Sitter Effective Theory (SdSET)~\cite{Cohen:2020php, Cohen:2021fzf} (see~\cite{Green:2022ovz} for review) has facilitated the systematic computation of corrections to the evolution equations of Stochastic Inflation (for alternative approaches, see~\cite{Burgess:2015ajz,Gorbenko:2019rza,Baumgart:2019clc,Mirbabayi:2019qtx,Mirbabayi:2020vyt,Baumgart:2020oby}).
In the presence of primordial non-Gaussianity (non-trivial interactions during inflation), perturbation theory for the probability distribution can be naturally organized as an Edgeworth series, such that the coefficients of the expansion are determined by the cumulants of the distribution.   However, such an expansion is expected to break down when computing observables that are sensitive to very rare configurations.

One natural hope is that this tail of the probability distribution can be captured by some kind resummation of the perturbative series. However, implementing such a resummation simply yields the Fokker-Planck equation for the distribution of cosmological fluctuations, with small corrections that are equivalent to performing a `local' Kramers–Moyal expansion of the underlying master equation~\cite{Cohen:2021fzf} (for an alternative approach, see~\cite{Celoria:2021vjw}). To assess the range of validity for these methods, a first step is to compute the corrections to these equations systematically, which allows one to explore the properties of the resulting probability distributions.  To this end, we computed the corrections to Stochastic Inflation from the leading effects of primordial non-Gaussianity in~\cite{Cohen:2021jbo}. By studying the phase transition to eternal inflation (an observable that is exponentially sensitive to the tails of the probability distribution), we observed that the resulting probability distribution was not under perturbative control.  While one might anticipate the tail is sensitive to non-perturbative effects, the precise origin and location of this breakdown should be calculable from the Effective Field Theory (EFT) point of view.

This paper will explain how the tails of these distributions are  determined by a new instanton-like\footnote{The saddles we will discuss are not instantons in the conventional sense.  However, they are occasionally referred to as instantons in the broader literature. } saddle point.  These new saddles have their own associated energy scale, which can invalidate the naive EFT expansion, where the IR scale is set by the Hubble scale $H$.  As we will quantify below, there are circumstances where these saddles are under control within the EFT description, and they simply reproduce the behavior of Stochastic Inflation.  However, when computing the probability for observables that are sensitive to sufficiently large deviations, the saddle lies beyond the EFT description and must be calculated by appealing to the full theory.

We will provide a framework for anticipating such a breakdown by recasting these questions in terms of random walks.
In fact, the appearance of the Fokker-Planck equation in inflationary cosmology suggests that the phenomenology of fluctuations in dS spacetime can be mapped onto the behavior of random walks. Specifically, we will show that random walks with independent and identically distributed (\iid)\ steps give rise to the same behavior as cosmological systems. Concretely, consider an \iid\ walk with zero mean that traverses a distance $X$ in a number of steps $N$.
For typical fluctuations ($X \propto \sqrt{N}$) the Central Limit Theorem (CLT) tells us that $X$ is Gaussian distributed, even if the individual steps are not. As we will review below, one can view the CLT as the result of a renormalization group flow to a Gaussian fixed point, with all non-Gaussianity being irrelevant -- typical fluctuations are insensitive to the microscopic details of the walk. On the other hand, large deviations ($X \propto N$) defy this general behavior and \textit{do} depend on the microscopic details. The fact that large deviations are not determined by universal long-distance behavior will have a precise analogy in cosmology, and will explain the breakdown of EFT for large fluctuations.
\begin{figure}
    \centering
    %
    %
    %
    %
    \includegraphics[scale=0.37]{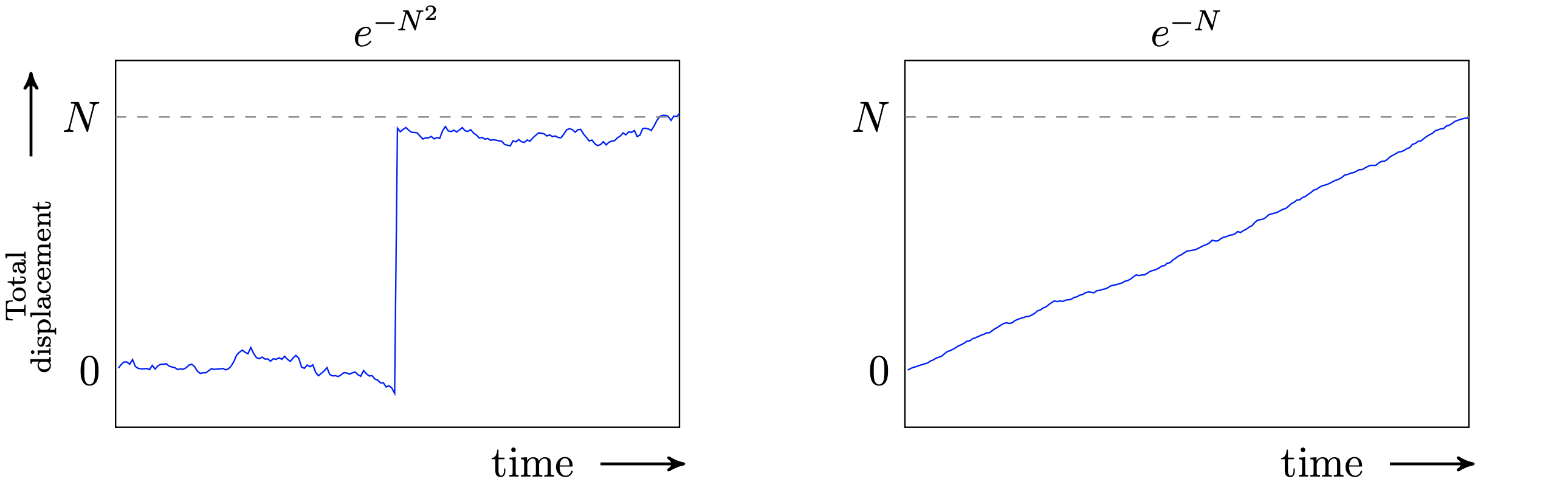}
    \caption{\label{fig:LargeFluctuations}
    Large deviations in a random walk. The walker makes one large jump of size $\L$ in the left figure, on top of many small positive and negative steps.  In contrast, the figure on the right depicts the walker covering the same distance $\L$ in a series of smaller steps that are mostly in the same direction. Although both of these are extremely unlikely, the one on the right is much more probable than the one on the left.  If we assume that the individual steps are sampled from a Gaussian distribution, the probability of making one large jump is $\sim e^{-\L^2}$, whereas the walker can get there using a series of smaller aligned steps with probability $\sim e^{-\L}$. Therefore, the latter dominates the tail of the coarse grained probability distribution. This shows that the most probable paths are described by small fluctuations around a single classical deterministic trajectory, which is associated with a novel saddle point solution.}
\end{figure}

A deeper understanding of such fluctuations can be gleaned from the Large Deviation Principle (LDP)~\cite{2008arXiv0804.2330V,2011arXiv1106.4146T}. Stated simply, the LDP is a scaling law of the form $P_N \simeq e^{-N I}$, where $P_N$ is a probability distribution parameterized by some large number $N$, and $I$ is a positive number called the rate function. In the context of random walks $I \sim O(1)$ for large fluctuations, which means $P_N \sim e^{-N}$ at the tail. As illustrated in \cref{fig:LargeFluctuations}, the dominant contributions to this tail come from walks that resemble a classical trajectory, which looks quite different from the usual zero mean walks. This is the new saddle that the CLT (or EFT) can fail to capture. The central goal of this work is to develop a precise map between the LDP framework and the physics of Stochastic Inflation.\footnote{Although to our knowledge, our paper is the first to make the connection between the LDP and the physics of inflation, the LDP has been previously applied to questions in observational cosmology~\cite{Bernardeau:2015khs, Uhlemann:2015npz, Reimberg:2017vnd, 2018sf2a.conf..257C,Ivanov:2018lcg, Barthelemy:2019ciu, Barthelemy:2020yva, Barthelemy:2020igw, Cataneo:2021xlx}.}

There is a vast literature on large deviations that extends well beyond \iid\ random walks, including applications to a number of physical problems in equilibrium and non-equilibrium statistical physics, \emph{e.g.}~see the review article~\cite{Touchette2009}.
For our purposes, the language of the LDP will be useful for two reasons: First, it makes precise the sense in which the physics leading to large deviations requires a departure from the usual long distance description in terms of the CLT and implies that one must appeal to the microscopic nature of the walk. Second, the rate functions $I$ are often calculable in terms of a novel saddle point approximation. Combining these two insights will allow us to characterize the regime of validity of Stochastic Inflation, highlighting the situations where it improves perturbation theory and why it ultimately breaks down. We will demonstrate this concretely in the context of eternal inflation and $\lambda \phi^4$ in a fixed de Sitter background. The LDP will explain the behavior of the probability distribution of scalar fluctuations calculated used the stochastic framework. We then extend this understanding to models that generate primordial black holes.

This paper is organized as follows.  We begin with a discussion of models that have a vanishing potential.  We first explore this scenario using random walks in \cref{sec:RandomWalks}. This allows us to show how the CLT emerges from a coarse graining procedure, and to both explain the LDP and apply it in a simple context.  Using this formalism, we can then understand the probability distribution for the fluctuations of the inflaton, which is the topic of \cref{sec:inflation}.  We then turn on a non-trivial potential for a random walk in \cref{sec:forces}, and explore the role of the LDP when computing the probability distribution for this example.  This is exactly the framework we need to understand the behavior of massless scalar field theory in a dS background in \cref{sec:scalarsindS}.  We then apply the same techniques to explore models whose goal is to generate primordial black holes in \cref{sec:PBHs}.  Finally, \cref{sec:conclusions} provides our conclusions and a discussion of future directions.

\section{Random Walks and the Renormalization Group}
\label{sec:RandomWalks}

Random walks offer a simple setting within which we can understand the conceptual details of the present work. We will review the Central Limit Theorem (CLT) and demonstrate how the long wavelength behavior of a wide class of walks fall into the universality class modeled by a Gaussian distribution \cite{McGreevyRG}. The failure of the CLT for large fluctuations is exactly analogous to the breakdown of EFT we discuss later in the paper. We can understand this regime better with the Large Deviation Principle (LDP), as illustrated with some simple examples worked out in detail below. This framework will eventually allow us to gain insight into the evolution of the scalar fluctuations of the inflaton from a new perspective.

\subsection{Central Limit Theorem}
\label{sec:CentralLimitThm}
\begin{figure}
    \centering
    %
    %
    %
    %
    \includegraphics[scale=0.37]{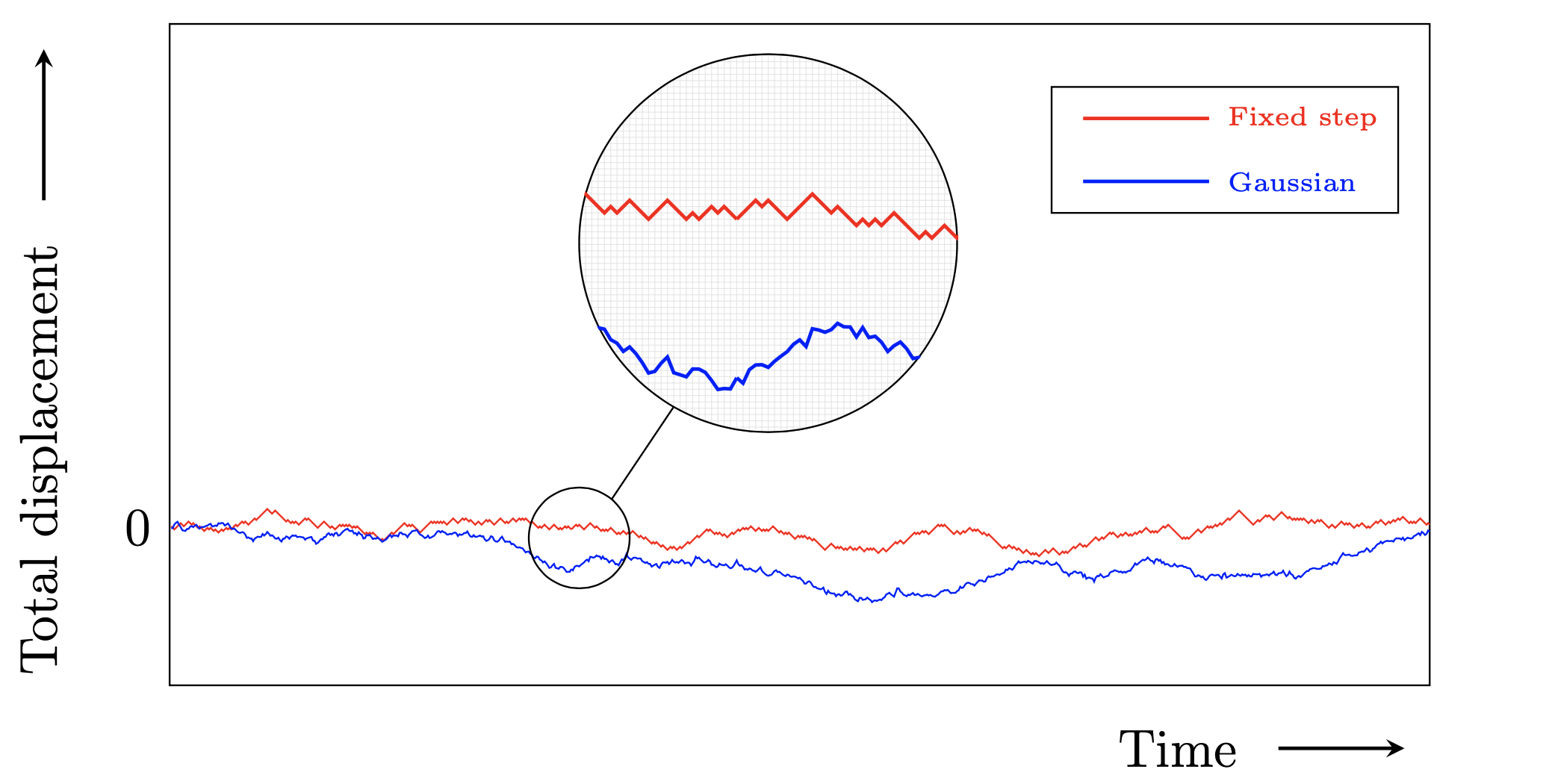}
    \caption{\label{fig:RWexamples} Random walks generated by a fixed step length distribution and a Gaussian distribution. If we zoom in, the difference in step size is evident from the shape of the trajectories, but at the macroscopic level it is difficult to tell which $p(x)$ generated each walk.}
\end{figure}

Consider a one dimensional \iid\ random walk. Starting from the origin, the walker takes one step per discrete time interval, each with a displacement $x$ chosen independently from some `microscopic' distribution\footnote{We use the word `distribution' to mean the probability density function for a continuous random variable, as well as the probability distribution for a discrete random variable. In some texts this word applies exclusively to the latter, but we make no such distinction in this work.} $p(x)$ with finite moments. To facilitate the discussion below, we will consider two specific examples:
\begin{subequations}
    \begin{align}
        p_f(x) &= \frac{1}{2} \delta (|x| - 1) \qquad \text{(Fixed step length distribution)} \ , \label{eq:PDFexamplesF}\\[5pt]
        p_g(x) &= \frac{1}{\sqrt{2 \pi}} e^{-x^2/2} \qquad \text{(Gaussian distribution)} \ . \label{eq:PDFexamplesG}%
    \end{align}
    \label{eq:PDFexamples}%
\end{subequations}%
Both of these distributions have mean $\langle x \rangle = 0$ and variance $\langle x^2 \rangle -\langle x \rangle^2 = 1$. A walker governed by $p_f(x)$ can take a step with either $x=1$ or $x=-1$ at each turn. The Gaussian walker's probability for each step is governed by $p_g(x)$, so it is able to take a step of any size, with the most probable values being $|x| \lesssim 1$. Examples of a typical random walk generated by $p_f(x)$ and $p_g(x)$ are given in Fig.\ \ref{fig:RWexamples}. If we zoom in on each walk we can make out the difference: it is evident that the red trajectory is borne of fixed size steps whereas the blue one is not. However, at the macroscopic level the two walks look like they could have been generated by the same $p(x)$, suggesting that the long wavelength behaviors of both walkers have something in common.

We can make this intuition precise by studying the net displacement of a random walker after $N$ steps, which we denote as
\begin{equation}
X = \sum_{i=1}^{N} x_i \ .
\end{equation}
If each step is sampled from $p(x)$, the probability of finding the walker at a distance $X$ from the origin after $N$ steps is
\begin{equation}
    P(X)
        = \int \prod_{i=1}^{N} \d x_i\, p(x_i)\,
        \delta \left( X - \sum_{i=1}^{N} x_i \right) \ .
    \label{eq:CoarseGrainedPDF}
\end{equation}
$P(X)$ can be evaluated by Fourier transforming the delta function,
\begin{align}
    \delta\bigg(X - \sum_{i=1}^{N} x_i\bigg) = \int \frac{\d k}{2\pi} e^{-i k\left(X - \sum_{i=1}^{N} x_i\right)} \ ,
\end{align}
to obtain
\begin{align}
    P(X) &= \int \frac{\d k}{2\pi}\, e^{-i k X} \int \prod_{i=1}^{N} \d x_i\, p(x_i)\s e^{i k x_i}\notag\\[5pt]
    &=
    \int \frac{\d k}{2 \pi}\, e^{-i k X}
        \left\langle e^{i k x} \right\rangle^N \ .
    \label{eq:PDFwithCharFn}
\end{align}
The second equality holds because we are assuming that the $x_i$ are \iid\ random variables, which means their expectation values are independent. The quantity
\begin{equation}
    \left\langle e^{i k x} \right\rangle
        = \int \d x \, p(x)\s e^{i k x} \ ,
    \label{eq:CharFn}
\end{equation}
is called the characteristic function of $p(x)$. For example $\left\langle e^{i k x} \right\rangle_g = e^{-k^2/2}$, if the steps are sampled from the Gaussian $p_g(x)$ in \cref{eq:PDFexamplesG}. Plugging this into \cref{eq:PDFwithCharFn} yields

\begin{equation}
    P_g(X) = (2 \pi N)^{-1/2} e^{-\frac{X^2}{2N}} \ ,\label{eq:PgX}
\end{equation}
which is an exact result valid for all $X$.

We can repeat this exercise with $p_f(x)$; after $N$ steps, the walker will be at $X$ with a probability
\begin{equation}
    P_f(X) = \frac{N!}{\left( \frac{N-X}{2} \right)! \left( \frac{N+X}{2} \right)!} \frac{1}{2^N} \ . \label{eq:FixedStepPDF}
\end{equation}
In the large $N$ limit, we can use Stirling's approximation $n! \xrightarrow{n \to \infty}n^n e^{-n}$ to write
\begin{equation}
    P_f(X) \simeq e^{-N I_f(X/N)} , \label{eq:PfX}
\end{equation}
with
\begin{equation}
    I_f(X/N) = \frac{1}{2} \left[ \left( 1-\frac{X}{N} \right) \ln\left( 1- \frac{X}{N} \right) + \left( 1+\frac{X}{N} \right) \ln\left( 1+\frac{X}{N} \right) \right] \ .
\end{equation}
If $X \ll N$,  we can Taylor expand $I_f(X)$ to find
\begin{equation}
    P_f(X) \simeq e^{-\frac{X^2}{2N} - \frac{X^4}{12 N^3} + \cdots} \ .
    \label{eq:GaussianPDFSmallX}
\end{equation}
So long as we restrict ourselves to $X \lesssim \sqrt{N}$, the first term in the exponent dominates, and we can approximate the distribution in \cref{eq:FixedStepPDF} as a Gaussian. In particular, $P_f(X)$ and $P_g(X)$ have the same behavior up to $X \sim \sqrt{N}$. A plot of $X$ over time can be obtained by downsampling the trajectories in Fig.\ \ref{fig:RWexamples} by a factor of $N$. From this perspective it is clear that $P(X)$ captures the long wavelength dynamics of the walk, which is the same at leading order for the fixed step length and the Gaussian walkers.

We can generalize these conclusions to a wide class of $p(x)$. Going back to \cref{eq:CharFn}, the Taylor expansion in $k$ of the logarithm of the characteristic function yields the cumulant expansion\footnote{The expression \cref{eq:CumulantGeneratingFunction} may be more familiar to some readers as the generating function for connected correlation functions, which are $\langle x^m \rangle_C$ here. In that language, the discussion above can be re-contextualized as 0+1 dimensional field theory.}
%
\begin{equation}
    \log \left\langle e^{i k x} \right\rangle
        = \sum_{m=1}^{\infty} \frac{(ik)^m}{m!} \langle x^m \rangle_C \ .
    \label{eq:CumulantGeneratingFunction}
\end{equation}
For instance, $\langle x \rangle_C \equiv \langle x \rangle$ and $\langle x^2 \rangle_C \equiv \langle x^2 \rangle - \langle x \rangle^2 = \sigma_0^2$ is the variance of $p(x)$. We can exponentiate the above expression to write \cref{eq:PDFwithCharFn} as
\begin{equation}
    P(X)
        = \int \frac{\d k}{2 \pi}
            e^{-i k X}
            \exp \left(ik N\langle x \rangle -\frac{1}{2} N k^2 \sigma_0^2 + \frac{i^3}{3!} N k^3 \langle x^3 \rangle_C + O(k^4)  \right) \ .
    \label{eq:CoarseGrainedPDF2}
\end{equation}
For observables where the Gaussian contribution dominates, the integrand has support for $k \lesssim \frac{1}{\sqrt{N} \sigma_0}$. Therefore, the $m^{\rm th}$ term in the cumulant expansion contributes to the exponent as
\begin{equation}
    N k^m \langle x^m \rangle_C \sim N^{1-m/2}\langle x^m \rangle_C
        \xrightarrow{N \to \infty} 0 \quad \text{for } m>2 \ ,
\label{eq:centrallimitthmrescaling}
\end{equation}
where $\langle x^m \rangle_C$ are independent of $N$ since the moments of the distribution $p(x)$ are finite constants.
Therefore, the first and second terms of the exponent in \cref{eq:CoarseGrainedPDF2} scale as $O(\sqrt{N})$ and $O(1)$ respectively and the remaining terms shrink as $O(1/\sqrt{N})$ or faster. This behavior can be made manifest by rescaling $k \to k/\sqrt{N}$ and $X \to \sqrt{N} X$ in \cref{eq:CoarseGrainedPDF2} so that
\begin{equation}
    P(X)
        \to \frac{1}{\sqrt{N}} \int \frac{\d k}{2 \pi}
            \exp \left(ik (\sqrt{N}\langle x \rangle-X) - \frac{1}{2} k^2 \sigma_0^2 + O\left( \frac{1}{\sqrt{N}} \right)  \right)
        \simeq e^{\frac{-(X - \sqrt{N}\langle x \rangle)^2}{2 \sigma_0^2}},
        \label{eq:CLT}
\end{equation}
which is a Gaussian probability distribution centered at $\sqrt{N} \langle x \rangle$ up to corrections that vanish for large $N$. This result is known as the \textit{Central Limit Theorem}.\footnote{The CLT is more general; it is not necessary for the steps to be independent\cite{Kardar_2007}. In that case \cref{eq:CLT} is still valid so long as the $m^{\rm th}$ cumulant in \cref{eq:CoarseGrainedPDF2} satisfies $\sum_{i_1, \cdots, i_m}^N \langle x_{i_1} \cdots x_{i_m} \rangle_C \ll O(N^{m/2})$.}

It is very useful to interpret this result in the language of Renormalization Group (RG) evolution as applied to EFTs. In this context, one identifies a power counting parameter which facilitates the use of dimensional analysis.  In the classic case of integrating out a heavy particle of mass $M$, the power counting is determined by the small dimensionless number $E/M$, where $E \ll M$ is the typical energy associated with the process of interest.  This allows one to organize the local operator expansion of the EFT into terms which are relevant (grow larger polynomially at lower energies), marginal (only evolve at most logarithmically), and irrelevant (grow smaller polynomially at lower energies).

We can see the same principles in action by viewing our random walk examples through the lens of RG.  If we organize the terms that appear in the exponent of \cref{eq:CLT} by how they scale with $N$, then we see that the mean is a relevant parameter, the variance is marginal, and the cumulants $\langle x^{m>2} \rangle_C$ are irrelevant.  Therefore, as $N\to \infty$, the distribution is localized about the mean, and the Gaussian distribution emerges as a universal fixed point of the RG evolution.  The interpretation is that coarse graining the distribution by zooming out (equivalently taking a large number of steps) erases any detailed memory of the microscopic distribution $p(x)$, beyond the gross features captured by its mean and standard deviation.  This is in exact analogy with EFTs, where only a small number of parameters contribute significantly to low energy observables, regardless of the detailed UV completion.

\subsection{Large Deviation Principle}
\label{sec:LDPOnRW}
However, this is not the whole story. Returning to the examples introduced in \cref{eq:PDFexamples}, let us consider the probability $P(X>N)$ of finding a random walker at a distance farther than $N$ from the origin, after $N$ steps. The walker taking fixed (unit) size steps has no hope of going beyond $N$ even if they were to take all $N$ steps in the same direction, and therefore $P_f(X>N) = 0$. However, the Gaussian walker, with the same mean and standard deviation, can have $P_g(X>N) \neq 0$.
Evidently, some information about the microscopic distribution $p(x)$ is encoded in the region $X \gtrsim N$, which we refer to as the \emph{tail} of $P(X)$.

For both examples above, it is extremely unlikely that the walker makes it to $X \sim N$ in $N$ steps, which implies $P(X \sim N)$ is very small. In order to probe the tail, we need to devise an observable that would be sensitive to such rare events. To this end, we can compute $\langle e^{\theta X} \rangle$ assuming $P_g(X)$ with $\theta > 0$. Naively, we might think that $\langle e^{\theta X} \rangle \sim e^{\theta O(\sqrt{N})}$ since $P_g(X)$ is dominated by $X \lesssim \sqrt{N}$. However, we can perform the following computation:
\begin{align}
\langle e^{\theta X} \rangle = \langle e^{\theta x} \rangle^N = e^{N\theta^2/2} \ ,
\end{align}
which in fact scales as $e^N$.
The explanation for the breakdown of the naive intuition is simply due to the fact that $e^{\theta X}$ takes on very large values with small probabilities, so that contributions from such values cannot be ignored \cite{10.1214/07-AOP348}. In other words, $\langle e^{\theta X} \rangle$ probes the tail of $P(X)$.

We now introduce a new random variable
\begin{equation}
    \X = \frac{1}{N} \sum_{i=1}^{N} x_i \equiv \frac{X}{N} \ , \label{eq:SampleMeanDefn}
\end{equation}
called the \emph{sample mean} of the \iid\ random variables $x_i$. Noting that the distributions transform under a change of variables as $P(\X) \d\X = P(X) \d X$, we may rewrite $P_g$ and $P_f$  for large $N$ as
\begin{align}
    P\big(\X\big) \simeq \exp\left(-N I(\X)\right) \ ,
    \label{eq:LDPforFandG}
\end{align}
where $I$ is a positive quantity called the \textit{rate function} that can be read off from \cref{eq:PgX} and \cref{eq:PfX}:
\begin{subequations}
    \begin{align}
       I_g(\X) &= \frac{\X^2}{2} \ , \\
       I_f(\X) &= \frac{1}{2}\big[\big(1-\X\big)\ln \big(1-\X\big) + \big(1+\X\big)\ln \big(1+\X\big)\big] \ .
        \label{eq:SampleMeanFixedStepLDP}%
    \end{align}%
    \label{eq:SampleMeanBoth}%
\end{subequations}%
A probability distribution $P_N$ that satisfies a scaling law of the form $P_N \simeq e^{-N I}$ is said to obey the \textit{Large Deviation Principle} (LDP).
\begin{figure}
  \centering
  %
  %
  %
  %
  %
  %
  \includegraphics[scale=0.35]{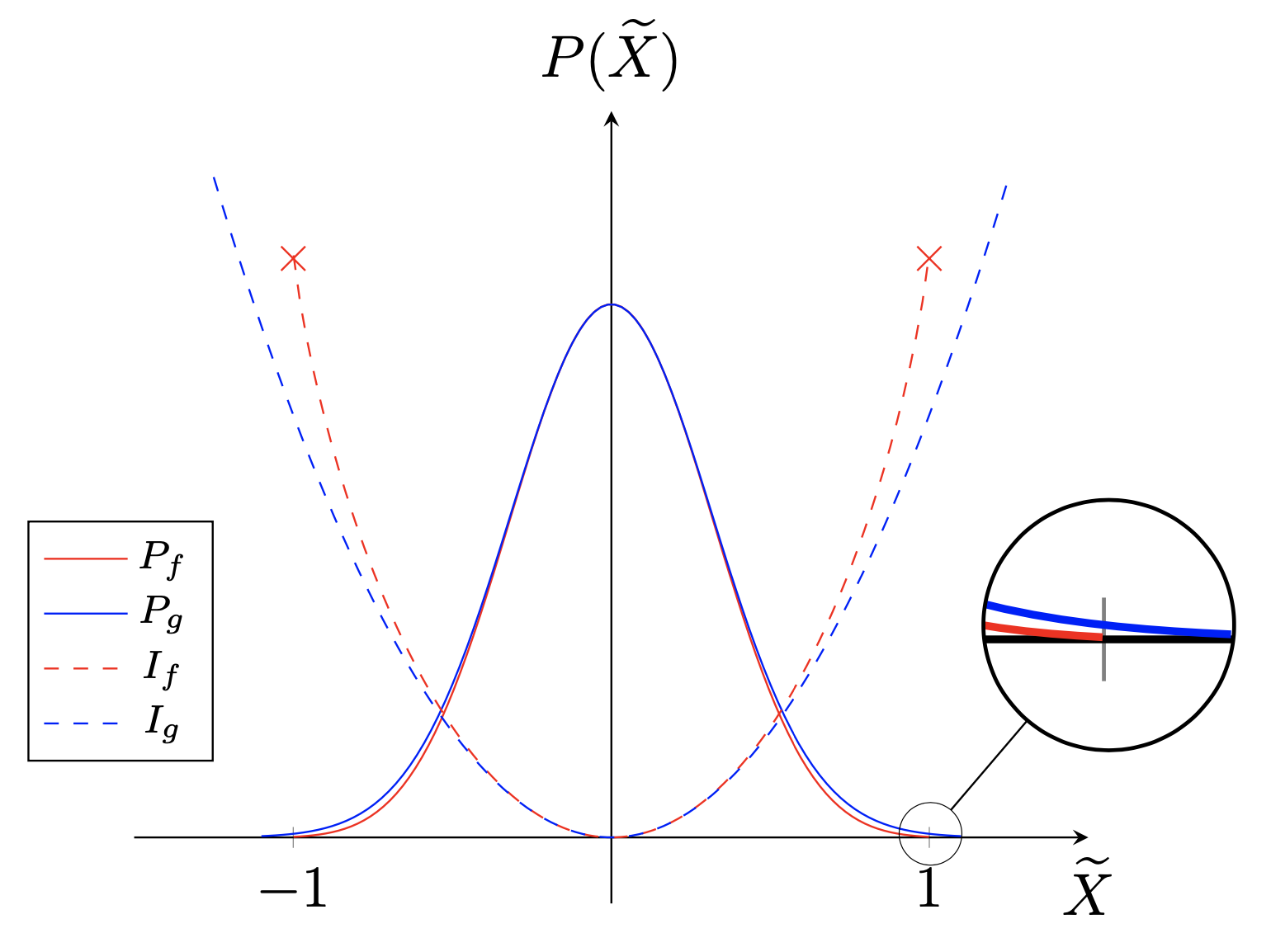}
  \caption{\label{fig:CLTvsLDP} Central Limit Theorem vs.\ Large Deviation Principle: The probability distributions of the sample mean $\X \equiv X/N$ for a fixed step length and Gaussian random walk for $N=10$, computed using the LDP. For values of $\X$ close to the mean both curves overlap, in accordance with the CLT. However, they differ for large deviations from the mean. In particular the tail of the distribution, shown magnified, reveals that $P_f(\X)$ vanishes at $\X = 1$ whereas $P_g(\X)$ does not. The dashed lines are the rate functions for each distribution.}
\end{figure}
The distributions of the sample mean described by \cref{eq:LDPforFandG} and Eqs.~(\ref{eq:SampleMeanBoth}) are examples of the LDP.

According to \textit{Cram\'er's theorem}, the distribution of a sample mean $\X$ of \iid\ random variables satisfies a LDP with a rate function $I(\X)$ given by
\begin{equation}
    I(\X) = \sup_\theta \left[ \theta \X - \lambda(\theta) \right] \ ,
    \label{eq:CramersTheorem}
\end{equation}
where $\lambda(\theta)$ is the cumulant generating function for each \iid\ variable\footnote{If the variables are not \iid\ we should use $\lambda(\theta) = \lim_{N\to\infty} \frac{1}{N} \ln \left[ \langle e^{\theta X} \rangle \right]$, the \textit{scaled} cumulant generating function, in \cref{eq:CramersTheorem}. This generalization is called the \textit{G\"artner-Ellis theorem}.}
\begin{equation}
    \lambda(\theta) \deq \ln \big\langle e^{\theta x} \big\rangle .
\end{equation}
To see why this is plausible, let us assume that LDP holds for some sample mean $\X$. Then, $P(\X) \simeq e^{-N I(\X)}$ and
\begin{equation}
    \big\langle e^{\theta X} \big\rangle = \big\langle e^{N \theta \X} \big\rangle
        \simeq \int \d\X e^{N\left(\theta\X - I(\X)\right)}
        \simeq e^{N \sup_\X \left[\theta\X - I(\X)\right]} .
\end{equation}
In the last step  we have used the saddle point approximation to compute the integral for large $N$. Noting that $\langle e^{\theta X} \rangle = \langle e^{\theta x} \rangle^N \equiv e^{N \lambda(\theta)}$ for \iid\ random variables, we have
\begin{equation}
    \lambda(\theta) = \sup_\X \left[\theta\X - I(\X)\right] .
    \label{eq:Cramer}
\end{equation}
Then the rate function given in \cref{eq:CramersTheorem} follows from a Legendre transform of this result.

As an example, we can work out the rate function $I_f(\X)$ for the sample mean of the fixed step length walk using Cr\'amer's theorem. Starting with $\langle e^{\theta x} \rangle = \cosh(\theta)$, we have
\begin{equation}
    I_f(\X) = \sup_\theta \left[ \theta \X - \ln \cosh(\theta) \right] .
    \label{eq:IFexample}
\end{equation}
The supremum can be obtained using ordinary calculus: taking the first derivative of the expression in brackets and setting it to zero gives $\theta_\text{max} = \tanh^{-1}(\X) = \frac{1}{2} (\ln(1+\X)-\ln(1-\X))$, which is where the r.h.s is maximum. Substituting this $\theta_\text{max}$ into \cref{eq:IFexample} and simplifying, we recover the rate function\footnote{$I_f(\X)$ is only defined for $\X \in [-1,1]$ on the real number line, which means the probability of finding $\X$ outside this interval is zero, as discussed earlier.} given in \cref{eq:SampleMeanFixedStepLDP}. Finally, notice that the rate function $I_f$ from \cref{eq:SampleMeanFixedStepLDP} is just the negative of the entropy for a binary random variable. We touch upon this fact in \cref{sec:EqDistr}. A more general discussion is given in \cite{Touchette2009}.

The CLT arises from the LDP when the rate function $I(\X)$ is convex and has a single global minimum (at say, $\X_0$). If this is the case, we can Taylor expand $I(\X)$ around $\X_0$ to obtain
\begin{equation}
    P(\X) \simeq e^{-\frac{1}{2} N I''(\X_0) (\X-\X_0)^2} \ .
\end{equation}
For \textit{small} deviations of $\X$ from $\X_0$, this quadratic expansion is a good approximation of $I(\X)$, and therefore the CLT provides the same information as the LDP. On the other hand \textit{large} deviations of $\X$ are those values at which the rate function deviates significantly from the quadratic approximation. The CLT does not correctly describe such large fluctuations, and we need to rely on the LDP instead.\footnote{That a random variable satisfies the CLT does not always imply that the existence of a rate function with a quadratic minimum, see example 3.4 of \cite{Touchette2009}.}

\section{Eternal Inflation}
\label{sec:inflation}

Having set up the general ideas of the LDP, we now turn to our first cosmological application.
The connection derives from the fact that scalar fluctuations during single-field inflation act locally like a 1-dimensional random walk around a classical trajectory. For a typical path, the end of inflation is determined by the classical evolution where the field distance changes linearly in time, $\Delta \phi_{\rm classical} =\dot \phi\s t$. However, it is possible for quantum fluctuations of the scalar field to work against the classical motion, giving rise to inflationary periods that last significantly longer than the classical expectation.  In fact, when the amplitude of fluctuations is large enough, it is known that inflation never ends everywhere~\cite{Linde:1986fd,Goncharov:1987ir} in the universe and instead gives rise to a infinite reheating volume~\cite{Creminelli:2008es}, also known as eternal inflation.  Remarkably, the fluctuations responsible for eternal inflation are necessarily examples of large deviations, as we will see in this section.

\subsection{Review of Stochastic Inflation}

The idea that inflation is essentially a random walk has a long history, starting from nearly the inception of the subject~\cite{Starobinsky:1986fx,Nambu:1987ef,Starobinsky:1994bd}.  The intuition follows from considering the freeze out of modes as they cross the horizon, at which point the quantum fluctuations of these modes begin to evolve classically.  In any small patch of the universe, the gradients of the field redshift away and the process is effectively a random walk.   (Of course globally, there are correlations across super horizon scales, which is the one of the main reasons we invoke inflation in the first place.) Specifically, as long as the parameters of the inflationary model change slowly in time, the fluctuations of each mode in any given patch of space follow the same distribution as a random walk with \iid\ variables, which is given the interpretation of \emph{noise} generated by the Hubble temperature associated with the horizon.  The distribution of fluctuations is also sensitive to the presence of a potential.  In the random walk language, this is the analog of a classical external force.  So the picture is a competition between the noise and this so-called \emph{drift}. This idea was formalized in the framework of Stochastic Inflation~\cite{Vilenkin:1983xq,Starobinsky:1986fx}, which is the statement that the probability distribution for the scalar fluctuations obeys
\beq\label{eq:old_stoc}
\frac{\partial}{\partial t} P(\phi,t)= \frac{H^3}{8 \pi^2} \frac{\partial^2}{\partial \phi^2} P(\phi, t)  + \frac{1}{3 H}  \frac{\partial}{\partial \phi} \big [V'(\phi) P(\phi,t ) \big]  \ .
\eeq
Despite its intuitive appeal, the derivation of Stochastic Inflation from quantum field theory, a full understanding of its domain of applicability, and a framework for computing corrections to the formalism had long been elusive.

These puzzles have recently been solved by interpreting Stochastic Inflation as arising from RG flow (or resumming logs) in quantum field theory~\cite{Burgess:2015ajz,Gorbenko:2019rza,Baumgart:2019clc,Mirbabayi:2019qtx,Cohen:2020php,Mirbabayi:2020vyt,Baumgart:2020oby,Cohen:2021fzf}. Concretely, by taking moments of \cref{eq:old_stoc}, one can relate mixing of operators under time-evolution to the stochastic equation. In single-field inflation, the fluctuations of $\phi$ can be rewritten in terms of the adiabatic metric fluctuation $\zeta$. However, $\zeta$ must respect the single-field consistency conditions~\cite{Maldacena:2002vr,Creminelli:2004yq,Creminelli:2012ed,Hinterbichler:2012nm}, which are the nonlinearly realized SO(4,1) symmetries that act on the metric leaving the gauge fixed.  For example, under the dilatation transformation in this group, $\zeta$ transforms as $\delta \zeta=-1-\vec{x} \cdot \vec{\partial}_{\vec{x}} \zeta$. The evolution of operators under (dynamical) RG must respect these symmetries and restricts the form of mixing to
\beq
\frac{\partial}{\partial\t} \zeta^N(\x,\t) =  \sum^N_{n\geq 2} \gamma_{n} \left(\!\!\begin{array}{c}
N \\
n
\end{array}\!\!\right) \zeta^{N-n}(\x,\t) \ ,
\label{eq:generalRG}
\eeq
where $\t = H\s t$, and $\gamma_n$ are the ``anomalous dimensions’’ which govern the composite operator mixing; the $\gamma_n$ are time-independent for scale invariant correlators. This implies the most general form of single field Stochastic Inflation is\footnote{In order to go from the dynamical RG for correlators to a Fokker-Planck equation for a probability distribution $P(\zeta,t)$, one simply identifies $\langle \zeta^n \rangle = \int \d \zeta P(\zeta, t) \zeta^n$.}
\beq\label{eq:general_FP}
\frac{\partial}{\partial\t} P(\zeta,\t) = \sum_{n\geq 2} (-1)^n \frac{\gamma_n}{n!} \frac{\partial^n}{\partial\zeta^n} P(\zeta, \t) \ .
\eeq

As discussed in \cite{Cohen:2021fzf},
we can view this as the expansion of a general Markovian process with transition amplitudes $W(\zeta | \zeta')$, such that
\begin{align}
\frac{\partial}{\partial \t} P(\zeta, \t) &= \int \d \zeta' \Big[ P(\zeta',\t)\s W(\zeta|\zeta') -P(\zeta, \t)\s W(\zeta'|\zeta) \Big] \notag \\[5pt]
&=\int \d \Delta \zeta \Big[ P(\zeta-\Delta \zeta,\t)\s \widetilde{W}(\Delta\zeta) -P(\zeta, \t)\s \widetilde{W}(\Delta \zeta) \Big]\,.
\label{eq:markovian}
\end{align}
Here we used the shift symmetry, $\zeta \to \zeta + c$, to write the transition amplitudes $W(\zeta | \zeta') = \widetilde W(\zeta-\zeta'\equiv \Delta \zeta)$. Taylor expanding $P(\zeta-\Delta\zeta,t)$ (\emph{a.k.a}~performing a Kramers–Moyal local expansion) reproduces \cref{eq:general_FP}, where
\beq
\gamma_n \equiv  \int \d \Delta \zeta\, \big(\!-\!\Delta \zeta\big)^n\,\widetilde W(\Delta\zeta)\, .
\eeq
In this sense, we see that $\gamma_{n>2}$ corresponds to non-Gaussian corrections to the transition amplitude, which is the same as the non-Gaussianity of the probability for a step in a \iid\ random walk.

The coefficients $\gamma_n$ are determined by computing the $n^\text{th}$ connected quantum field theory correlator through
\beq
\langle \zeta^n(\x,\t) \rangle \supset \gamma_n \log aH \ .
\eeq
Explicit calculation shows that $\gamma_1=0$, which is a restatement of the conservation of $\zeta$ outside the horizon. The quadratic term $\gamma_2$ is determined by the variance
\beq
\langle \zeta^2(\x=0) \rangle = \Delta_\zeta \int \frac{\d^3 k}{(2\pi)^3} \frac{1}{2 k^3} = \frac{\Delta_\zeta}{4 \pi^2}\log \frac{a H}{K_\text{IR}} \ ,
\eeq
where $\Delta_\zeta = H^4 /  (2 f_\pi^4)$ sets the amplitude of the power spectrum for $\zeta$, and we evaluated this integral by introducing a hard UV cutoff $\Lambda = a H$ and an IR cutoff $K_\text{IR}$.\footnote{While this is the correct result, one might be concerned that this hard cutoff breaks spacetime symmetry.  For a discussion of a dimensional regularization-like regulator that preserves the symmetries, see~\cite{Cohen:2020php, Premkumar:2021mlz}.}
Comparing \cref{eq:generalRG} to \cref{eq:general_FP}, we see that this term generates the noise term that appeared in the original formulation of Stochastic Inflation, \cref{eq:old_stoc}.

From the point of view of the quantum field theory correlators and the resultant RG evolution, computing higher order corrections is completely straightforward.  Applying this approach to the EFT of inflation~\cite{Creminelli:2006xe,Cheung:2007st}, the first non-trivial correction to the stochastic framework in single-field inflation was found in~\cite{Cohen:2021jbo}.
Since $\zeta$ is derivatively coupled, we can generically generate $\gamma_n$ by introducing an interaction of the form $\dot \zeta^n /\Lambda^{4-n}$, where $\Lambda = f_\pi c_s$ is the approximate UV cutoff\footnote{When $c_s \to 1$, there are additional factors of $(1-c_s^2)$ so that $\Lambda \to \infty$ as $c_s \to 1$ in slow-roll models.} of the EFT of inflation when $c_s \ll 1$.  Then by dimensional analysis,
\beq\label{eq:c_n}
\gamma_n = c_n \Delta_\zeta^{n/2} \left( \frac{H^2}{\Lambda^2}\right)^{2-n} \ ,
 \eeq
with $c_n = {O}(1)$. This leads to the naive expectation that perturbation theory should hold as long as one is working in the parameter space where $H \ll \Lambda$.  As was emphasized in \cite{Cohen:2021jbo}, this is only true when the observable of interest is insensitive to the tail of the probability distribution.  In the language of the LDP, these tails are dominated by a new saddle point.  The energy scale associated with the LDP saddle can be significantly larger that $\Lambda$, signaling that one is sensitive to the details of the UV completion, as in the case of the random walk examples studied above.

\subsection{Central Limit Theorem as a Resummation}

We would now like to solve for the time evolution of $P(\zeta,t)$, assuming $\zeta= \zeta_0$ at $t=0$. This will tell us the probability of different possible values of $\zeta$, which should resemble a random walk.  If the theory is Gaussian, so that $\gamma_{n>2} =0$, then we are solving the heat equation
\beq\label{eq:gaussian_FP}
\frac{\partial}{\partial\t} P(\zeta,\t) =  \frac{\sigma^2}{2} \frac{\partial^2}{\partial\zeta^2} P(\zeta, \t) \ ,
\eeq
where $\sigma^2 \equiv \gamma_2 = \Delta_\zeta /(4\pi^2)$ is the variance. The solution to this equation is a Gaussian
\beq
\label{eq:gaussian_sol}
 P_\text{G}(\zeta,\t;\zeta_0) = \frac{1}{\sqrt{2\pi \sigma^2 \t} } e^{- (\zeta-\zeta_0)^2/(2 \sigma^2 \t)} \ .
\eeq
We can use the Gaussian solution to construct the solution to \cref{eq:general_FP} with general $\gamma_n$, in terms of derivatives of $ P_\text{G}$
\beq
P(\zeta,\t;\zeta_0) = \exp\left(\sum_{n>2} (-1)^n \, \frac{\gamma_n \t}{n!} \frac{\partial^n}{\partial \zeta^n} \right) P_\text{G}(\zeta,\t;\zeta_0)  \ .
\label{eq:Pzetasol}
\eeq
Using this general form, we will show that the physics of the random walk is reproduced by the solutions to this equation.  First, let us consider the behavior around the peak of the Gaussian solution where $(\zeta-\zeta_0)^2 \simeq \sigma^2 \t$. If we expand the full solution near the peak, we notice that as $\t \to \infty$,
\begin{align}
    \frac{\gamma_n \t}{n!} \frac{\partial^n}{\partial \zeta^n} P_\text{G}(\zeta,\t;\zeta_0)
        &\simeq \frac{\gamma_n \t}{n!} \frac{(-1)^n (\zeta-\zeta_0)^n}{(\sigma^2 \t)^n} P_\text{G}(\zeta,\t;\zeta_0) \label{eq:ZetaScaling} \\[0.5em]
        &= {O}(\gamma_n \t^{1- n/2} \sigma^{-n})  P_\text{G}(\zeta,\t;\zeta_0) \ .
\end{align}
If we associated $\gamma_n$ with the $n^\text{th}$ cumulant of a random walk, and $\t \to N$ is the number of steps, then the suppression of these terms precisely matches our expectations from the CLT as $\t \to \infty$, see \cref{eq:centrallimitthmrescaling}.

Note that this implies that $\zeta \sim 1$ is under control for suitably large $\t$. In contrast, perturbative calculations of the probability distribution using the Edgeworth series
\begin{align}
P(\zeta) \simeq  \exp \left(- \int \frac{\d^3 k}{(2\pi)^3} \frac{\zeta(\k) \zeta(-\k)}{2 P(k)} +\int \frac{\d^3 k_1 \d^3 k_2}{(2\pi)^6} \frac{B(k_1,k_2, k_3)  \zeta(\k_1) \zeta(\k_2) \zeta(\k_3)}{6 P(k_1)P(k_2) P(k_3)}+\ldots \right)  \ , \notag
\end{align}
where $\k_3= -\k_1-\k_2$, breaks down for much smaller values of $\zeta$. This shows how Stochastic Inflation improves the behavior of perturbation theory by resumming the individual modes into a single random walk.

\subsection{Large Deviations and the EFT of Inflation}
\label{sec:LDPEFTInf}
Now let us consider the tail of the $P(\zeta,\t)$ distribution, where $\zeta =\alpha \t$ for some constant $\alpha$ in the limit $\t \to \infty$. The region $\alpha \geq 1$ corresponds to the regime of eternal inflation, as the random fluctuations conspire to prevent the end of inflation, even in the $\t \to \infty$ limit. The transition at $\alpha =1$ is where the quantum fluctuations exactly cancel the classical evolution of the background field. Note that because the distance is linear in $\t$, rather than $\sqrt{\t}$, we are considering a large deviation for the probability distribution of $\zeta$.

It is straightforward to see that for these large deviations the CLT fails to calculate dominant contribution to the tail, just as it did for the \iid\ random walk (see \cref{fig:RWexamples}). Plugging $\zeta = \alpha \t$ into \cref{eq:ZetaScaling}, we have
\beq \label{eq:order}
\frac{\gamma_n \t}{n!} \frac{\partial^n}{\partial \zeta^n} P_\text{G}(\zeta,\t;\zeta_0)  = {O}(\gamma_n \t \alpha^n \sigma^{-2n})  P_\text{G}(\zeta,\t;\zeta_0) \ .
\eeq
For $\alpha = {O}(1)$, there is no suppression of the higher-order terms. Concretely, the entire series in $\gamma_n$ will break down for sufficiently large $\alpha$. For $\alpha =1$, this series will break down when $\Lambda < f_\pi$, even though this parameter space is consistent with condition that the EFT of inflation is weakly coupled at horizon crossing, $\Lambda > H$~\cite{Cohen:2021jbo}.

The breakdown of Stochastic Inflation is a precise reflection of what we found in our analysis of large deviations for random walks.  To see this more clearly, we can write the solution to Stochastic Inflation in terms of the Fourier transform of \cref{eq:Pzetasol}
\beq\label{eq:FT}
P(\zeta,\t) = \int_{-\infty}^{\infty} \d k\, e^{-i k \zeta} \rho(k,\t) =  C \int_{-\infty}^{\infty} \d k \exp\left(-i k \zeta -k^2 \frac{\sigma^2}{2} \t -i k^3 \frac{\gamma_{3}}{3!} \t +\ldots \right) \ .
\eeq
We recognize this as precisely the result for a random walk we described above, see~\cref{eq:CoarseGrainedPDF2}. At the same time, we can identify
\beq
\rho(k,\t) = C \exp\left( -k^2 \frac{\sigma^2}{2} \t + \sum_{n>2} (i k)^n \frac{\gamma_{n}}{n!} \t \right) \equiv \langle \exp\left(i k \zeta\right) \rangle \ ,
\eeq
so that when $\zeta = \t \alpha$, we have $\rho(k,\t) \to e^{\t \lambda(\theta=i k)}$. Assuming $\t \gg 1$, we can calculate the integral over $k$ in \cref{eq:FT} using the method of steepest descents:
\beq
P(\zeta=\alpha \t, \t) \simeq \exp \left(- \t I(\alpha) \right) \ ,
\eeq
where
\beq\label{eq:FT_rate}
I(\alpha) = \left(-i \alpha k_\star(\alpha) -k_\star^2(\alpha) \frac{\sigma^2}{2} +\sum_{n>2} \left(i k_\star(\alpha) \right)^n \frac{\gamma_{n}}{n!} \right) \ .
\eeq
The integrand has been expanded around $k=  k_\star(\alpha)+\delta k$ defined by a (complex) value of $k$ that is an extremum of the argument of the exponential,
\beq
\left(i \sum_{n\geq 2}  \left(i k_\star(\alpha) \right)^n \frac{\gamma_{n+1}}{n!} \right)  -k_\star \sigma^2 = i \alpha\ .
\eeq
We see that using the method of steepest descents to calculate the inverse Fourier transform is equivalent to using Cram\'er's theorem, \cref{eq:CramersTheorem}. Furthermore, for large $\alpha$ the Gaussian solution, $k_\star =(-i) \alpha /\sigma^2$, is a far from the true saddle as all the terms in the $k_\star^n \gamma_n$ expansion will become equally important. This is, of course, the Fourier transform of the result in \cref{eq:order}.

When we applied the LDP to random walks in \cref{sec:LDPOnRW}, it was clear that we become sensitive to the microphysics. We would like to understand this breakdown purely in terms of the EFT of Inflation. Concretely, the expansion in $\gamma_n$ is under control at horizon crossing, which is the physical energy scale where the fluctuations are produced.  A natural guess is that $\zeta \propto \t$ behaves like a classical solution with $\dot \zeta = H$ or $\dot \phi = f_\pi^2$, where $\phi$ is the inflaton. To make sense of this, we can rewrite the evolution of $\zeta$ in terms of a Langevin equation,
\beq
\frac{\d}{\d\t} \zeta(\t)  = \xi(\t) \ ,
\eeq
where $\xi(\t)$ is a random variable that models a noise source.  Assuming that the noise is Gaussian, we have
\beq
\langle \xi(\t) \xi(\t') \rangle  = \sigma^2 \delta(\t-\t') \ .
\eeq
The probability of finding $\zeta = \zeta_f$ at $\t =\t_f$ given the initial condition $\zeta(\t=0) =0$ is then
\beq
P(\zeta_f) = \int {\cal D}\zeta(\t) \exp\left(-\int_0^{\t_f} \d\t \, \frac{1}{2 \sigma^2} \left(\frac{\d}{\d\t} \zeta(\t)\right)^2  \right) \ .
\eeq
For large deviations, $\zeta = \alpha \t$, and the probability is determined by the saddle point $\frac{\d^2}{\d\t^2} \zeta = 0$ or $\frac{\d}{\d\t} \zeta = \alpha$, so that
\beq
P(\zeta_f) \simeq \exp\left(- \t \frac{\alpha^2}{2\sigma^2} \right) \ .
\eeq
When $\alpha = {O}(1)$ (\emph{i.e.}~not ${O}(\t^{-1})$) this satisfies the LDP.

The key observation is that the probability of this large deviation is determined by a classical solution where $\frac{\d}{\d\t} \zeta = \alpha$ (see \cref{fig:LargeFluctuations}).  Translating this into the canonically normalized field, $\zeta_c = \zeta f_\pi^2/H$, this is the condition that
\beq
\frac{\d}{\d t} \zeta_c = \alpha f_\pi^2 \ .
\eeq
The EFT of inflation is defined in terms of an expansion in $\dot \zeta_c / \Lambda^2$ and therefore when $\dot \zeta_c > \Lambda$, we cannot define these classical solutions within the EFT.  Concretely, we can modify the Langevin equation with nonlinear terms
\beq
\left(1+\sum_{n>1} c_n \left(\frac{f_\pi^2}{\Lambda^2} \frac{\d\zeta_c}{\d\t} \right)^n \right) \frac{\d}{\d\t} \zeta_c(\t)  = \xi(\t) \ ,
\eeq
where $c_n ={O}(1)$ by the definition of $\gamma_n$ in terms of $\Lambda$ in \cref{eq:c_n}.  We can now calculate the probability distribution as before,
\beq
P(\zeta_f) = \int {\cal D}\zeta_c(\t) \exp\left(-\int_0^{\t_f} \d\t \, \frac{\left(\dot \zeta_c(\t)(1+ \sum_{n>1} c_n \dot \zeta_c^n f_\pi^{2n}/\Lambda^{2n})\right)^2 }{2\sigma^2} \right) \ .
\eeq
The saddle point is still $\ddot \zeta_c =0$, but we can see that the probability distribution
\beq
P(\zeta_f) \simeq \exp\left(-\t\frac{\alpha^2}{2\sigma^2}\left(1+\sum_{n>1} c_n \frac{\alpha^n f_\pi^{2n}}{\Lambda^{2n}}\right)^2\right)
\eeq
becomes ill-defined when $\alpha f_\pi^2 > \Lambda^2$. It is also noteworthy that the breakdown of EFT in this specific example is not associated with the breakdown of Markovian dynamics, as higher time derivatives vanish around the classical solution (see the discussion in \cref{sec:markovian}). For single field inflation, it is a breakdown of the EFT of Inflation itself, rather than SdSET, that is responsible for the ill-defined probability distribution for sufficiently large deviations.

\section{Random Walks with External Forces}
\label{sec:forces}

In the previous sections, we focused on the application of the LDP to random walks with no external deterministic forces. We argued that the late time behavior for the typical fluctuations of these systems could be determined by RG evolution.  This analysis yielded the CLT, such that the resultant probability distribution was a Gaussian with zero mean.

In this section, we will study the physics of a random walk that is driven by an external deterministic force.  This is easiest to understand in the case of a constant force, which is equivalent to an \iid\ random walk where the average over steps is non-zero, $\langle x \rangle \neq 0$.  Then from \cref{eq:CLT}, we see that the term proportional to the non-zero mean scales as $\sqrt{N}$, and so this term grows as we take $N\to \infty$.  In the RG language, this implies that an external deterministic force has the effect of introducing a relevant deformation into the theory.

\subsection{Equilibrium Distributions}
\label{sec:EqDistr}

In the presence of confining forces, such as a potential with a local minimum, we might expect to see ergodic behavior, such that the probability for being at a given location at a fixed time approaches a time-independent (equilibrium) distribution.  Such behavior is also consistent with our expectations from thermodynamics for large numbers of confined particles. In fact, it turns out that the equilibrium distribution for this thermodynamic system is itself a quantity that is calculable using the LDP. If we imagine that a walk of length $N$ reaches equilibrium, then the probability of finding the particle at location $y$ during the walk at a sufficiently large number of steps, $1 \ll n  \leq N$, should simply follow the equilibrium distribution
\beq
P(x_n=y) = P_{\rm eq}(y) \ .
\eeq
Now suppose that we calculate quantities averaged over the entire walk, such as $e^{\theta X}$ where $X=\sum_i x_i$. Since we are averaging over the locations of each step, assuming we have reached equilibrium, we have
\beq
\bigg\langle \exp\bigg(\theta \sum_{i=1}^N x_i\bigg) \bigg\rangle \simeq \left(\int \d y\, e^{\theta y} P_{\rm eq}(y)\right)^N = e^{N \lambda(\theta)} \ ,
\eeq
where $\lambda(\theta)$ is determined from the equilibrium distribution for a single step in the walk. This is a qualitative argument that can be formalized in terms of the eigenvalues of the transition amplitudes. The rate function for the walk $X= \sum_{i=1}^N x_i$ again follows from \cref{eq:Cramer} and \cref{eq:SampleMeanDefn},
\beq
\lambda(\theta) = \sup_\X \left[\theta\X - I(\X)\right] \ .
\eeq
Importantly, in the limit $N\to \infty$, the probability for any average quantity, $S= \sum_{i=1}^n f(x_i)$, is just the $N^\text{th}$ power of finding one particle with $f(x)=S/N$.

The appearance of an equilibrium quantity in the LDP calculation is not a special feature of random walks, but is common to most statistical mechanics problems~\cite{Touchette2009}. In a precise sense, rate functions of the LDP are proportional to the thermodynamic free energies for large numbers of particles.  The overall power of $N$ in the probability is just the familiar relationship between extensive and intensive thermodynamic quantities. In fact, this connection was already present when we calculate the rate function for the discrete walk in \cref{eq:SampleMeanFixedStepLDP}, which is (minus) the entropy associated with a binary random variable.

This perspective helps explain why the LDP can be used calculate the equilibrium probability distribution. This can be made concrete\footnote{This equation is formally meaningless, as the random variable is not differentiable. One can formalize these results using the It\^o or Stratonovich prescriptions. This subtlety will not play an important role in our discussion, see~e.g.~\cite{Pinol:2020cdp} for more details.  } in terms of a Langevin equation:
\beq\label{eq:eom}
\dot x(t) =  f\big(x(t),t\big) + \xi(t) \ ,
\eeq
where again $\xi(t)$ is a random variable that accounts for noise, and now $f\big(x(t),t\big)$ is an external deterministic force. If we assume the noise is Gaussian, then $\xi(t)$ obeys
\beq\label{eq:xi2pt}
\big\langle \xi(t)\s \xi(t') \big\rangle  = \sigma^2 \delta(t-t') \ .
\eeq
In this case, the probability of a walk $x(t)$ is
\begin{align}
P\big[x(t)\big] &=C \exp\left(-\frac{1}{2\s\sigma^2} \int_0^T \d t\s \big(\xi(t) \big)^2 \right) \notag\\[5pt]
&=C \exp\left(-\frac{1}{2\s\sigma^2} \int_0^T \d t\s \big(\dot x(t) - f\big(x(t),t\big)\big)^2 \right)  \ ,
\end{align}
where we used the equations of motion given in \cref{eq:eom}. One can confirm the first line by taking functional derivatives with respect to $\xi(t)$ to reproduce the two-point correlator in \cref{eq:xi2pt}. For making future contact with cosmology, we will assume the external force is due to a potential such that
\beq
f\big(x(t),t\big) \to - V'(x(t)) \ ,
\eeq
where $V' \equiv \partial_x V(x)$.

Now suppose we are given $x(0)=0$, and we want to integrate over all possible paths to find the probability that $x(T) = L$. We can solve this problem using the method of steepest descents. First, we must find the maximum likelihood path, which is the same as finding the classical saddle for the `effective action'
\beq
I(L) = \frac{1}{2}\int_0^T \d t \s \big(\dot x +V'\big)^2 = \frac{1}{2}\int_0^T \d t \s \big(\dot x^2 + V'{}^2\big) + \big[V(x)\big]\big|_0^T \ .
\label{eq:ILintermediate}
\eeq
From the equations of motion, we have
\beq\label{eq:solution_wV}
\ddot x = V'' V' \quad\to\quad \frac{\d}{\d t} \dot x^2 = \dot x \frac{\d}{\d x} \big(V'^2\big) \quad\to\quad \dot x^2  = V'^2 \ .
\eeq
Using this result, we can write
\beq
\frac{1}{2} \int_0^T \d t\s \big(\dot x^2 + V'{}^2\big)  =\frac{1}{2}  \int_0^T \d t \, 2\s\dot x\s V' =  \int_0^T  \d t \,  \frac{\d}{\d t} V =  \big[V(x)\big]\big|_0^T\ .
\eeq
Including the total-derivative term from \cref{eq:ILintermediate}, the probability of find $x(T)=L$ is given in terms of
\beq
I(L) = 2 \big(V(L) - V(0)\big) \qquad\text{with}\qquad P(L) \simeq \exp(- I(L) / \sigma^2 ) \ .
\eeq
Here our determination of the probability, $P(L)$, is a LDP result in the sense that $I(L)$ is a rate function
\beq
\lim_{\sigma \to 0} \sigma^2 \log P(L) = - I(L) \ .
\eeq
In the same sense as for the path integral, it is the paths near the classical solution that yield the dominant probability, while the contribution from the fluctuations about the classical path determine the sub-leading terms in the expansion with respect to $\sigma$.

This result matches the equilibrium probability distribution we derive from the Fokker-Planck equation
by setting $\d P/\d t = 0$:
\beq
\frac{\d}{\d x} \big(V' P(x) \big) + \frac{\sigma^2}{2} \frac{\d^2}{\d x^2}P(x) = 0 \ .
\eeq
Integrating twice with respect to $x$ gives
\beq
\log \frac{P(L)}{P(0)} = -\frac{2}{\sigma^2}\big(V(L) -V(0) \big) \ .
\label{eq:genericPeq}
\eeq
We see that the LDP reproduces the equilibrium distribution that is predicted by the Fokker-Planck equation.

\subsection{Markovian Evolution}
\label{sec:markovian}

Within this framework, we can also easily understand the role of Markovian evolution when assessing the validity of the calculation of the equilibrium solution. Markovian refers to a class of theories where the next time step is fully determined by the state of the system at the previous time step.  In terms of differential equations, Markovian evolution therefore is equivalent to the statement that we have a first-order (in time) equation of motion.  If there were higher derivatives, then one would need to know about the state of the velocity field along with the state of the system itself to determine the next step in the evolution~\cite{VANKAMPEN198569}.

We can therefore model non-Markovian evolution by adding a small acceleration term to our equations of motion given in \cref{eq:eom}:
\beq
\vartheta \ddot x(t) + \dot x(t) =  -V'\big(x(t)\big) + \xi(t) \ ,
\eeq
so that the dynamics are Markovian in the $\vartheta  \to 0$ limit. Repeating our previous calculation, we find an effective action
\begin{align}
I(L) &=\frac{1}{2}  \int_0^T \d t \big(  \vartheta\s\ddot x(t) +\dot x + V'\big)^2 \notag\\[5pt]
&= \int_0^T \d t \big(\vartheta^2\s \ddot x^2 +\dot x^2\big(1-\vartheta V''\big)+ V'{}^2\big) + \big[V(x)+ 2\s\vartheta\s \dot x\s V'(x)+\vartheta\s \dot x\big]\big|_0^T \ , \label{eq:theta_corr}
\end{align}
so that $P(L) = \exp\big(-I(L)/\sigma^2\big)$.  There are two new non-Markovian terms that contribute to the action, which are small corrections when
\begin{subequations}
\begin{align}
\vartheta^2 \ddot x^2 = \vartheta^2 (V'' V')^2 &\ll \dot x^2  = V'^2 \label{eq:nonMarkoviana} \\[4pt]
\vartheta V'' &\ll 1 \ ,
\label{eq:nonMarkovianb}
\end{align}%
\label{eq:nonMarkovian}%
\end{subequations}%
where the first and second lines correspond the first and second $\vartheta$-dependent terms in \cref{eq:theta_corr}, and the equalities in \cref{eq:nonMarkoviana} are due to the \cref{eq:solution_wV}. Notice that both terms remain small when we impose the condition $\vartheta V'' \ll 1$ everywhere along the path.  If we enforce Eqs.~(\ref{eq:nonMarkovian}), then the effective action is first order in time, and hence the evolution is Markovian.

\section{Light Scalar Fields in de Sitter}
\label{sec:scalarsindS}

Light scalar fields in de Sitter with non-trivial potentials present an additional complication beyond single-field inflation. The stochastic framework applied to these models is known to give rise to a non-Gaussian equilibrium probability distribution, acting as a kind of non-trivial fixed point of the dynamical RG. This presents a vastly different situation, compared to single-field inflation, where interactions are negligible for typical fluctuations due to the CLT.  This section will show how the dynamics of these models can be mapped onto the language of random walks with external forces that we developed in the previous section.

\subsection{Effective Potentials and Markovian Dynamics}

Stochastic Inflation provides a compelling framework with which to understand the dynamics of light scalar fields in dS. In its original form, it describes the probability distribution for an interacting scalar in dS, via the Fokker-Planck equation
\beq
\frac{\partial}{\partial t} P(\phi,t)= \frac{H^3}{8 \pi^2} \frac{\partial^2}{\partial \phi^2} P(\phi, t)  + \frac{1}{3 H}  \frac{\partial}{\partial \phi} \big [V'(\phi) P(\phi,t ) \big] \ .
\label{eq:FPeqStocInf}
\eeq
The equilibrium probability distribution is given by
\beq
P_{\rm eq}(\phi) \simeq \exp\left(-\frac{8 \pi^2 V(\phi)}{3H^4} \right) \ .
\label{eq:Peq}
\eeq
If we take $V(\phi) = \lambda \phi^4/4!$, the most likely field values are $|\phi| \lesssim H \lambda^{-1/4}$. In light of the discussion of non-Gaussian noise in \cref{sec:inflation}, one would naturally wonder about the regime of validity and corrections to this formula.

It is useful to discuss Stochastic Inflation and its corrections in the context of SdSET. The relationship between the variables of SdSET and scalar field theory in dS can be understood from the SdSET ansatz for a free massless scalar:
\beq\label{eq:ansatz}
\phi(\k, \t)  = H \left( \vp(\k, \t) + \vm(\k, \t)  [a(\t) H]^{-3} \right) \ .
\eeq
Here we have rewritten the scalar in terms of the two power law solutions as $k\to 0$, where $\vp$ is the constant (or growing) mode and $\vm$ is the decaying mode.

In SdSET, Stochastic Inflation is a consequence of Callan-Symanzik-like equations for the dynamical RG of the $\vp^n$ operators.  This information can be rewritten as a master equation for the probability distribution $P(\vp,\t)$, which at lowest order reproduces \cref{eq:FPeqStocInf}, while also containing an infinite series of corrections:
\begin{align}\label{eq:stocastic_expand}
    \frac{\partial}{\partial \t} P(\vp,\t) &=\frac{1}{3}  \frac{\partial}{\partial\vp} \left [V'_{\rm eff} (\vp) P(\vp,\t ) \right] \notag\\[5pt]
    &\hspace{14pt}+ \frac{\partial^2}{\partial \vp^2}  \left[ \sum_{m=0}^\infty \frac{ b_m}{2!} \vp^{2m} P(\vp, \t) \right]+  \frac{\partial^3}{\partial \vp^3}  \left(\vp \sum_{m=0}^\infty \frac{d_m}{3!} \vp^{2m} P(\vp, \t) \right) \notag\\[5pt]
&\hspace{14pt} +  \frac{\partial^4}{\partial \vp^4}  \left( \sum_{m=0}^\infty \frac{ e_m}{4!} \vp^{2m} P(\vp, \t)  \right) + \ldots \ .
\end{align}
For the UV example of a massless scalar with $V(\phi) = \lambda \phi^4/4!$, the leading corrections (as defined below) were calculated in~\cite{Cohen:2021fzf}, resulting in
\begin{align}
\hspace{-10pt}\frac{\partial}{\partial \t} P(\bvp,\t)
    &= \frac{1}{3}  \frac{\partial}{\partial\bvp} \big [V'_{\rm eff}(\bvp) P(\bvp,\t ) \big]
    +\frac{1}{8\pi^2}\frac{\partial^2}{\partial \bvp^2} P(\bvp, \t) \notag \\[5pt]
   &\hspace{14pt}+ \frac{\lambda_{\rm eff}}{1152\pi^2}  \frac{\partial^3}{\partial \bvp^3}  \big(\bvp   P(b\bvp, \t) \big) \ ,
    \label{eq:FPatNNLO}
\end{align}
with
\begin{align}
    V'_{\rm eff} &= \frac{\lambda_{\rm eff}}{3!} \bigg( \bvp^3 + \frac{\lambda_{\rm eff}}{18}\bvp^5 + \frac{\lambda_{\rm eff}^2}{162} \bvp^7 +\,...\, \bigg)\ ,
    \label{eq:VpEff}
\end{align}
where we redefined
\beq
\lambda_{\rm eff}= \lambda + 18 b_2  \qquad\text{and}\qquad \vp = \bvp + \frac{b_1}{6 b_0} \bvp^3
\eeq
to remove the scheme-dependent corrections $b_1 = {O}(\lambda)$ and $b_2 ={O}(\lambda^2)$.

First, let us establish in what sense these are small corrections to the original Fokker-Planck equation. If we ignore the corrections to
the evolution \cref{eq:FPatNNLO} and the potential \cref{eq:VpEff}, so that $V_{\rm eff}(\vp) = \lambda \vp^4/4!$, then the equilibrium solution is the same as in \cref{eq:genericPeq}:
\beq
P^{\rm LO}_{\rm eq}(\vp) = \exp\left(-\frac{2 V_{\rm eff}(\vp)}{3 \sigma^2} \right) =\exp\left(-\frac{\pi^2 \lambda \vp^4}{9} \right) \ ,
\eeq
where we substituted $\sigma^2 = \gamma_2 = (4\pi)^{-1}$ and $V' \to V'_{\rm eff} /3$ to match \cref{eq:FPatNNLO}.  Notice that the typical fluctuations reside in the region $| \vp|\lesssim \lambda^{-1/4}$. We can determine the scaling behavior of the solutions using $\vp \sim \lambda^{-1/4}$ such that the corrections to $V_{\rm eff}(\vp)$ are ${O}(\lambda_{\rm eff}^{1/2})$ and ${O}(\lambda_{\rm eff})$, which we will call next-to leading order (NLO) and next-to-next-to leading order (NNLO) respectively.  The cubic-derivative term, on the second line of \cref{eq:FPatNNLO}, is similarly NNLO. By the same $\lambda$-scaling argument, the equilibrium solution can be written as $P_{\rm eq} = C P_{\rm LO}(\vp) P_{\rm NLO}(\vp) P_{\rm NNLO}(\vp)$ with
\begin{subequations}
\begin{align}
P_{\rm LO} &= \exp\left(- \frac{\pi^2}{9} \lambda_{\rm eff} \vp^4 \right)\\[4pt]
P_{\rm NLO} &= \exp\left(- \frac{\pi^2}{243} \lambda_{\rm eff}^2 \vp^6 \right) \\[4pt]
P_{\rm NNLO} &=  \exp\left( \frac{5}{10368} \lambda_{\rm eff}^2 \vp^4 - \frac{17\pi^2}{46656} \lambda_{\rm eff}^3 \vp^8 \right)\,.
\end{align}%
\label{eq:Peqs}%
\end{subequations}%
The terms in $P_\text{NNLO}$ come with different powers of $\vp$ but have the same $\lambda_{\rm eff}$ counting for typical fluctuations.  Importantly, the second term, which is ${O}(\vp^8)$, receives contributions from both the change to $V_{\rm eff}(\vp)$ and the higher derivative term.

Given that our UV theory only has a marginal coupling, $\lambda \phi^4$, it is not obvious that there should be a breakdown of the stochastic framework akin to what happened for the EFT of Inflation in \cref{sec:inflation}.  Furthermore, we saw in \cref{sec:forces} that the equilibrium distribution is itself a result of the LDP, and therefore it is not a given that the framework could break down.

However, as we saw in \cref{sec:markovian}, a critical assumption for the validity of the stochastic framework is that the evolution is Markovian.  Therefore, the stochastic description can fail when the acceleration terms become important. The condition for non-Markovian terms to be negligible was given in Eqs.~(\ref{eq:nonMarkovian}) above.  In the language of SdSET, non-Markovian evolution would arise from nontrivial mixing between $\vp$ and $\vm$, defined in \cref{eq:ansatz}.

Evaluating these conditions requires that we identify the parameter $\vartheta$. For a light scalar field $\phi$ in dS, the equations of motion in the limit that $\k \to 0$ are
\beq
\ddot \phi(t) + 3 H \dot \phi(t) = -V'(\phi) \ .
\eeq
In terms of dimensionless time $\t = H t$, we would therefore\footnote{The non-Markovian form of Stochastic Inflation has been use in the literature, where $\vartheta =1/3$ as stated, \textit{e.g.}~\cite{Woodard:2005cw,Achucarro:2021pdh,Burgess:2022nwu}. It would be interesting to understand these non-Markovian terms in SdSET where the higher derivative corrections have been computed. } expect $\vartheta \simeq 1/3$. Assuming that $\vartheta^{-1} = O(1)$, \cref{eq:nonMarkovianb} implies that we should worry that the evolution becomes non-Markovian when
\beq
V''_{\rm eff} = {O}(1) \quad\to\quad \lambda \vp^2 ={O}(1) \ .
\eeq
Using the explicit form of the corrections provided in \cref{eq:Peqs}, we see that this is precisely where our expansion in powers of $\lambda$ breaks down:
\beq
P_{\rm LO}\left(\vp\sim \lambda^{-1/2}\right) \simeq P_{\rm NLO}\left(\vp\sim \lambda^{-1/2}\, \right) \simeq P_{\rm NNLO}\left(\vp\sim \lambda^{-1/2}\, \right) \sim \exp\big(-\lambda^{-1}\big)\ .
\eeq
Similarly, this is the scale where the infinite series of corrections to the effective potential in \cref{eq:VpEff} become equally important. It is therefore natural to conclude that the breakdown in our perturbative expansion in $\lambda$ when $\vp > \lambda^{-1/2}$ is due to the failure of the Markovian assumption.

\subsection{Light Scalars in de Sitter with Derivative Interactions}

We argued in \cref{sec:LDPEFTInf} that the breakdown of Stochastic Inflation for large fluctuations in single-field inflation was associated with the breakdown of the EFT of inflation. In this section, we will expore if a similar breakdown occurs for the equilibrium distribution of a scalar field $\phi$ described by an EFT with higher-derivative interactions in addition to a potential.

We will take our action for our scalar to be an EFT that includes arbitrarily high powers of derivatives\footnote{This is not to be confused with SdSET, which only describes the long wavelength modes in dS.} to take the form
\beq
S = \int \d^4 x\, \sqrt{-g} \left(\frac{1}{2}\partial_\mu \phi \partial_\mu \phi - V(\phi) + \sum_{n>1} \frac{y_n}{\Lambda^{4(n-1)}} ( \partial_\mu \phi \partial_\mu \phi )^n + \ldots \right) \ ,
\eeq
where $\Lambda$ is the UV cutoff of the EFT and the $\ldots$ include operators with more than one derivative per field. This description is under control in de Sitter when $\Lambda \gg H$. We will again take $V(\phi) = \lambda \phi^4 /4!$ such that $\phi$ is massless and its growing mode $\vp$ will evolve at zeroth order in $y_n$ according to $\lambda$-corrected equations of Stochastic Inflation given in \cref{eq:FPatNNLO,eq:VpEff}.

The impact of the higher-derivative couplings $y_n$ on Stochastic Inflation is nearly identical to the corrections in single-field inflation. The leading corrections in $y_n$ survive the $\lambda \to 0$ limit and therefore can be determined independent of the potential. In this limit, $\phi$ has a shift symmetry $\phi \to \phi+c$. Repeating the argument used for single field inflation in \cref{sec:inflation}, one finds corrections
\beq\label{eq:scalar_EFT}
\frac{\partial}{\partial\t} P(\vp,\t) \supset \sum_{n> 1} \frac{\gamma_n}{(2n)!} \frac{\partial^{2n}}{\partial\vp^{2n}} P(\vp, \t) \ ,
\eeq
where
\beq
\gamma_n \propto y_n\left(\frac{H}{\Lambda} \right)^{4(n-1)} \ ,
\eeq
at leading order in $y_n$.

The presence of higher-derivative terms in the scalar EFT introduces an infinite series of derivatives in the effective Fokker-Planck equation. For this to be under control, we expect that the equilibrium solution with $\gamma_{n>2} =0$ should be corrected by an expansion in powers of a small parameter.  If we write $P(\vp,\t) = P^\text{eq}_{\rm LO}(\vp) Q(\vp,\t)$, then in the limit $\vp \gg \lambda^{-1/4}$, \cref{eq:scalar_EFT} becomes
\beq
\frac{\partial}{\partial \t} \log Q \simeq \sum_{n>2} \frac{\gamma_n}{(2n)!} \left(-\frac{2  V_{\rm eff}'(\vp)}{3 \sigma^2} \right)^{2n} \ ,
\eeq
where we used \cref{eq:Peq} for $P_\text{LO}^\text{eq}(\vp)$.
For $y_n = {O}(1)$, this series is under control when
\beq
\frac{8 \pi^2 H^2}{3 \Lambda^2} V'_{\rm eff}(\vp) \ll 1 \ .
\eeq
Taking $V_{\rm eff} \simeq \lambda \vp^4$, this tells us that the equilibrium solution is under control for $H\vp \lesssim \Lambda^2 /
(\lambda H)$, which is parametrically larger than $\Lambda$. We can make sense of the regime of validity of this result using \cref{eq:solution_wV} to relate
\beq
V'_{\rm eff}(\vp) =| \dvp | \ ,
\eeq
along the classical trajectory. Now using $\phi \simeq H \vp$ and $\t = H t$, we can rewrite this condition for the expansion to be under control:
\beq
\frac{|\dot \phi|}{\Lambda^2} \ll 1  \ .
\eeq
We see that the breakdown is precisely where we would expect from the derivative expansion of the microscopic theory. Critically, the derivation of this breakdown required knowledge of the LDP to see that the equilibrium distribution could be derived from a classical saddle, in the regime where $\dvp$ was larger than allowed by the UV cutoff of the EFT of inflation.

\section{Models of Primordial Black Hole Generation}
\label{sec:PBHs}
The impact of rare fluctuations is particularly important for models of primordial black hole (PBH) generation. The PBHs are formed from  order-one fluctuations directly in the primordial distribution and, as such, the fluctuations are exponentially unlikely for scale-invariant Gaussian random fields. Models for the generation of PBHs therefore exploit breakdowns of both scale-invariance and Gaussianity, see \emph{e.g.}~\cite{Carr:2020xqk,Bird:2022wvk} for recent reviews.

The typical approach to estimating the abundance of PBHs follows from the critical collapse model~\cite{Carr:2020xqk}. In the conventional description, one takes a smoothed density field,
\beq\label{eq:smoothing}
\delta_R(\x) = \int \d^3 k\s e^{-i \k \cdot \x\,} W(k R)\s \delta(\k) \ ,
\eeq
where $W(\tilde k)$ is a filter that removes power on scales $\tilde k \gg 1$, and $R$ is some distance scale. The critical collapse model assumes that any region where $\delta_R > \delta_{\rm cr}$, for some constant threshold $\delta_{\rm cr} = {O}(1)$, will form a collapsed object with a total mass determined by the size of the region $R$.

Within this framework, the abundance of primordial black holes is determined by the probability of finding $\delta_R > \delta_{\rm cr} = {O}(1)$, where the precise value of $\delta_{\rm cr}$ is model-dependent. This threshold can be also be written as a critical value of $\zeta_R(\x)$~\cite{Kopp:2010sh,Harada:2013epa}, $\zeta_{\rm cr}$, defined as in \cref{eq:smoothing} such that the probability of finding $\zeta_R(\x) > \zeta_{\rm cr}$ determines that production of PBHs. For concreteness, a value of $\zeta_{\rm cr} = 0.1-0.2$ arises in some analytic collapse models in radiation domination~\cite{Harada:2013epa}. In comparison with the LDP, note that the relevant time-scale for the random  walk is the number of e-folds of inflation after horizon crossing for a model with $k R \sim 1$, or $N_e(R) = \log R a(t_{\rm end}) H$, where $t_{\rm end}$ is the time when inflation ends.  For scales $N_e(R) = {O}(10)$, fluctuations above the $\zeta_{\rm cr}$ threshold would correspond to $\alpha \gtrsim 10^{-2}$ using the parameterization of large deviations described in \cref{sec:LDPEFTInf}. In models of inflation consistent with observations, these values of $\alpha$ may still lie outside the domain of the EFT of inflation.

Non-Gaussian tails arise in a variety of contexts, included single- and multi-field inflation. In light of the connection between tails of distributions and the LDP explored in this paper, we would like to understand when such large non-Gaussian contributions can be calculated reliably given only an effective description at horizon crossing.  We will argue that framing these questions in the language of RG for a random walk provides useful intuition.

\subsection{Non-Gaussian Tails}

We have explained how the stochastic approach to inflation translates the problem of finding the distribution of scalar fluctuations onto characterizing the behavior of a random walk. The CLT tells us that the Gaussian probability distribution is a fixed point of the conventional random walk. Just as with RG flows in quantum field theories, we can classify the deformations that could produce a non-Gaussian tail, in analogy with \cref{sec:CentralLimitThm}, into three types: relevant, marginal, or irrelevant.
\vskip 5pt
\noindent {\bf Relevant:} A non-zero mean, \emph{e.g.}~due to a deterministic force, takes us away from the Gaussian fixed point of the CLT. For inflation, this corresponds to a potential $V(\phi)$ such that the equilibrium probability distribution takes the form of \cref{eq:genericPeq}, namely
\beq
P(\phi) = e^{- 2 V(\phi) / 2\sigma^2} \ .
\eeq
If the potential includes any operators other than a mass term, this distribution is non-Gaussian. Yet, since it is due to the presence of the unique relevant deformation, a large deviation from Gaussianity does not indicate a breakdown of the effective description.

In practice, non-trivial production rates for PBHs require some more complicated and possibly non-analytic potential $V(\phi)$. Some UV models may motivate particular non-perturbative shapes for $V(\phi)$, but in practice the formation of PBHs has mostly been explored using phenomenological models for the inflationary potential, see~\emph{e.g.}~\cite{Franciolini:2018vbk,Vennin:2020kng,Cai:2022erk}.
\vskip 5pt
\noindent {\bf Marginal:} A marginal deformation of a random walk corresponds to changing the covariance matrix that governs the steps in the walk.  In the context of inflation, this means changing the amplitude of scalar fluctuations, $P_\zeta(k)$, or mixing the inflaton with additional fields.  The former is a common strategy but only leads to enhanced Gaussian tails.  Mixing with additional fields can give rise to non-Gaussian tails in a variety of ways.

A canonical example of models that use mixing are the curvaton or modulated reheating scenarios, where the late time adiabatic mode is determined by a spectator field $\chi$, so that
\beq\label{eq:zeta_mix}
\zeta \simeq F(\chi) \to \langle e^{J \zeta}  \rangle = \langle e^{J F(\chi)} \rangle \ ,
\eeq
for some model-dependent function $F$. As a concrete example, suppose our spectator field has a potential $V(\chi) = \lambda \phi^4$ so that at leading order we have
\beq
P(\chi)=\exp\left(-\frac{\pi^2 \lambda \chi^4}{9} \right) \ .
\eeq
Now suppose that by some process after inflation, the adiabatic mode is determined by $\zeta = \kappa \chi^3$ for some constant $\kappa$.  By the change of variables (integrating over $\chi$ subject to the mixing with $\zeta$), we have
\beq
P(\zeta)=\exp\left(-\frac{\pi^2 \lambda |\zeta|^{4/3}}{9 \kappa^{4/3}} \right) \ .
\eeq
In this way, we can produce non-analytic behavior in the tail from otherwise local interactions. Of course, this assumes that the functions $V(\chi)$ and $F(\chi)$ are known exactly, when in fact they are themselves expansions in $\chi$. For a given model, one must check the self-consistency of truncating these expansions, including the corrections to $V_{\rm eff}(\chi)$ discussed in \cref{sec:scalarsindS}.

The derivation of \cref{eq:zeta_mix}, while a trivial restatement of the mixing, has an important interpretation in the context of the LDP.  In the LDP literature, this change of variables is known as the correspondence principle, which says that when $\zeta = F(\chi)$ and the rate function for $\chi$ is known, then the rate function for $\zeta$ is given by
\beq
I(\zeta) = \inf_{\chi:F(\chi)=\zeta} \, \tilde I_\chi(\chi) \ ,
\eeq
where $\tilde I(\chi)$ is the rate function that determines the large deviations of the $\chi$ field.

These types of probability distributions can arise from interactions that mix the adiabatic and isocurvature modes during or after inflation~\cite{Bond:2009xx,Chen:2018uul,Franciolini:2018vbk,Panagopoulos:2019ail,Palma:2019lpt,Palma:2020ejf,Achucarro:2021pdh,Hooshangi:2021ubn,Pi:2022ysn,Gow:2022jfb}. The probability distributions found in these examples match the discussion given here as they are well-described by the LDP. In SdSET, one can remove these types of mixing interactions via a field redefinition, which effectively introduces a transformation of the form \cref{eq:zeta_mix} on the observable fluctuations.

\vskip 5pt
\noindent {\bf Irrelevant:}
Non-Gaussian noise is an irrelevant perturbation of a random walk. The CLT ensures that even a highly non-Gaussian probability distribution will produce a Gaussian distribution for the total distance of the walk. We saw that this is not true for large deviations, which lie outside the regime of the CLT, but also require exact knowledge of the non-Gaussian probability distribution.

It is tempting to use non-Gaussian statistics for quantum fluctuations as a mechanism to produce PBHs. However, as discussed in \cref{sec:inflation} (and~\cite{Cohen:2021jbo}), when the non-Gaussian terms in Stochastic Inflation become important, both the stochastic framework and the EFT of inflation are breaking down. In principle one can use the LDP techniques to calculate the rate of functions within the microscopic model of inflation; however, given that the stochastic framework does not apply to the microscopic theory, one must go beyond the classical probabilistic description we have used here.

\subsection{Relation to Factorial Enhancement}

The increased probability distribution for large fluctuations has been tied to the factorial enhancement of higher-order correlators in a number of examples \cite{Flauger:2016idt,Munchmeyer:2019wlh,Panagopoulos:2019ail,Panagopoulos:2020sxp}. Concretely, if one is calculating the correlators of some field $\chi$ as a perturbative expansion in a parameter $\kappa \ll 1$, then the $M$-point correlators are factorial enhanced when~\cite{Panagopoulos:2020sxp}
\beq
\langle \chi^M ... \rangle \propto \kappa^m M! \propto M! \, \langle \chi^m \rangle^{M/m} \ .
\eeq
This scaling with $M$ is significant as it implies that there is more information encoded in the largest $M$-point functions.

The factorial enhancement can be understood more directly from the generating function $W[J]$, which is defined by the partition function with a source $J$ as\footnote{The notation in this section follows \cite{Panagopoulos:2020sxp}. It can be related to the symbols introduced in \cref{sec:LDPOnRW} via the following map: $J \to \theta, Z[J] \to \langle e^{\theta X} \rangle, W[J] \to - N \lambda[\theta], V(\chi) \to NI(\X)$.}
\beq
Z[J] = e^{-W[J]} = \left\langle \exp \left(-\int \d^d x\, J(\x) \chi(\x) \right) \right\rangle  \ .
\eeq
The $M^\text{th}$ connected correlator is then determined from the generating functional as
\beq
\langle \chi(\x_1) .. \chi(\x_M) \rangle_C = \frac{\delta^M}{\delta J(\x_1) .. \delta J(\x_M)} W[J]|_{J=0} \ .
\eeq
If we can locally expand $W[J]$ as
\beq\label{eq:W_exp}
W[J] = \sum_m a_m J^m \ ,
\eeq
then the correlators will be factorially enhanced unless $a_M \propto 1/M!$. This condition implies that the expansion in \cref{eq:W_exp} converges everywhere in the complex plane and therefore $W[J]$ is an entire function. It was confirmed by explicit calculation in~\cite{Panagopoulos:2020sxp} that $W[J]$ has a logarithmic  branch cut\footnote{As discussed in~\cite{Panagopoulos:2020sxp}, the existence of a point in the complex plane where $Z[J] =0$ is sufficient to demonstrate the $W[J]$ is not entire.} for $V(\chi) \propto |\chi|^p$ when $p>2$. In this precise sense, the non-Gaussian tails that are calculable via Stochastic Inflation also imply a factorial enhancement of the large-$M$-point correlators.

For the models described by \cref{eq:zeta_mix}, the generating functional was calculated in~\cite{Panagopoulos:2020sxp} using the method of steepest descents. It is noteworthy that $W[J]$ is the same quantity as $\lambda[\theta] \to W[J]$ that appears in Cram\'er's theorem, see \cref{eq:Cramer}. Concretely, $W[J = ik]$ is simply the Fourier transform of $P(\chi)$ so that
\beq\label{eq:W_to_P}
P(\chi) = \int_{-\infty}^{ \infty} \d k\, e^{-i k \chi}\exp(- W[i k]) \ .
\eeq
As we saw in \cref{eq:FT_rate}, when the LDP holds, the inverse Fourier transform can be calculated from the method of steepest descents and reproduces Cram\'er's theorem. Given that $W[J]$ is itself calculable by the non-trivial saddle, \cref{eq:W_to_P} can be interpreted as a Legendre transform from the generating functional to the rate function $W[J] \to I(\chi)$. The appearance of the Legendre transform in calculating this rate function using the LDP is equivalent to the role of the Legendre transform in relating free energies in statistical mechanics.

The role of the Fourier transform in relating the language of Stochastic Inflation and rare fluctuations also appears in the tail expansion of the probability distribution reviewed in \emph{e.g.}~\cite{Vennin:2020kng}. In that case, one is Fourier transforming the time variable, rather than the field $\chi$, but the $\delta N$-formalism ultimately relates the two at the end of inflation. In principle, the techniques of the LDP should also apply directly to the tail expansion and might offer insights into the regime of control of those calculations.

Finally, the above discussion is also related to the factorial enhancement of scattering amplitudes at high multiplicity~\cite{Cornwall:1990hh,Goldberg:1990qk,Brown:1992ay,Gorsky:1993ix,Libanov:1994ug,Son:1995wz}. In that context, semi-classical solutions have also proven to be important and are closely related to the semi-classical calculation of $W[J]$ described above. It is likely there is a deeper connection to the LDP, as we have seen in the case of cosmological correlators.

\section{Conclusions}
\label{sec:conclusions}

In this paper, we demonstrated that the Large Deviation Principle can be used to diagnose the validity of the underlying Effective Field Theory expansion being used to derive the evolution equations of Stochastic Inflation.  We showed how to interpret the dynamical Renormalization Group equations that derive Stochastic Inflation as coarse-graining a random walk.  When the potential is essentially zero, for example in the case of the inflaton,  we argued that this procedure leads to a Gaussian distribution as a consequence of the Central Limit Theorem.  In this case, EFT expectations hold and everything is under perturbative control.  However, if one asks questions that are sensitive to the tails of the probability distribution, then the LDP tells us that a new saddle point of the action dominates, and making a reliable prediction requires knowledge of the EFT to all orders (or equivalently one must appeal to the UV completion).  We then showed how the LDP applies for models with a non-trivial potential, and again explored the regime of EFT validity.  Finally, we showed how the LDP could be used to diagnose the validity of models that were introduced with the goal of yielding a non-trivial production of primordial black holes.

There are many important future directions to explore.  It would be of great interest to apply the LDP to compute the stochastic evolution equations in UV complete examples in such a way that the impact on the tails of the distributions was completely under control.  This would provide a test case analog of the random walk examples that were presented in \cref{sec:RandomWalks} above.  It would also be interesting to explore other applications of the LDP in cosmology and quantum field theory.  For example, the appearance of additional saddles describing the tail of the distribution is reminiscent of scattering amplitudes with high-multiplicity~\cite{Cornwall:1990hh,Goldberg:1990qk,Brown:1992ay,Gorsky:1993ix,Libanov:1994ug,Son:1995wz} and the large charge expansion of conformal field theories~\cite{Hellerman:2015nra, Badel:2019oxl, Grassi:2019txd, Badel:2019khk, Gaume:2020bmp}. These are natural settings where one might expect the LDP to play a role, and it would be exciting to make this precise.  We anticipate that having connected the validity of Stochastic Inflation to the LDP will yield many new insights into the nature of quantum field theory, both in dS spacetime and beyond.

\paragraph{Acknowledgements} \hskip 5pt We are grateful to Cliff Burgess, Paolo Creminelli, Paolo Glorioso, John McGreevy, Mehrdad Mirbabayi, Eva Silverstein, and Raman Sundrum.
TC is supported by the US Department of Energy under Grant~\mbox{DE-SC0011640}.
DG is supported by the US~Department of Energy under Grant~\mbox{DE-SC0009919}.
AP was supported in part by the Kavli Institute for Cosmological Physics at the University of Chicago through an endowment from the Kavli Foundation.
This work was completed at the Aspen Center for Physics, which is supported by the NSF grant PHY-1607611.

\phantomsection
\addcontentsline{toc}{section}{References}
\small
\bibliographystyle{utphys}
\bibliography{dSRefs}

\end{document}